\documentclass[10pt]{article}

\usepackage{amssymb}
\usepackage{amsmath}
\usepackage{amsthm}
\usepackage{amsfonts}
\usepackage{xcolor}
\usepackage{mathbbol}
\usepackage{bm}
\usepackage{url}
\usepackage{ dsfont }
\usepackage{times}
\usepackage{graphicx}
\usepackage{bm}

\usepackage{dsfont}
\usepackage[plain,noend]{algorithm2e}
\usepackage{titling}

\usepackage[numbers]{natbib}

\usepackage[utf8]{inputenc}
\usepackage{setspace}
\usepackage[margin=1in]{geometry}
\title{Regression in Tensor Product Spaces by the Method of Sieves}
\date{\vspace{-5ex}}
\author{Tianyu Zhang and Noah Simon}
\usepackage[page]{appendix}
\newtheorem{theorem}{Theorem}[section]
\newtheorem{lemma}[theorem]{Lemma}
\newtheorem{corollary}[theorem]{Corollary}
\newtheorem{definition}[theorem]{Definition}
\newtheorem{example}[theorem]{Example}
\newtheorem{proposition}[theorem]{Proposition}
\begin{document}
\maketitle
\vspace{-10mm}
\begin{center}
Department of Biostatistics, University of Washington\\ 1400 NE Campus Parkway, Seattle 98195, U.S.A.
\end{center}

\begin{abstract}
Estimation of a conditional mean (linking a set of features to an outcome of interest) is a fundamental statistical task. While there is an appeal to flexible nonparametric procedures, effective estimation in many classical nonparametric function spaces (e.g., multivariate Sobolev spaces) can be prohibitively difficult -- both statistically and computationally -- especially when the number of features is large. In this paper, we present (penalized) sieve estimators for regression in nonparametric tensor product spaces: These spaces are more amenable to multivariate regression, and allow us to, in-part, avoid the curse of dimensionality. Our estimators can be easily applied to multivariate nonparametric problems and have appealing statistical and computational properties. Moreover, they can effectively leverage additional structures such as feature sparsity. In this manuscript, we give theoretical guarantees, indicating that the predictive performance of our estimators scale favorably in dimension. In addition, we also present numerical examples to compare the finite-sample performance of the proposed estimators with several popular machine learning methods. 
\end{abstract}


\section{Introduction}
\label{section:introduction}

Understanding the relationship between an outcome of interest and a set of predictive features is an important topic across domains of scientific research. To this end, one often needs to estimate an underlying ``predictive" function, e.g., the conditional mean function, that best relates the features and the outcome using the available noisy observations. During the past two decades, there has been extensive research focusing on nonparametric learning methods that only require the outcome to vary ``smoothly" with the features.

One challenge of applying nonparametric methods in multivariate problem is ``the curse of dimensionality" \cite{richard1961adaptive}. Briefly speaking, we need exponentially more data to achieve the same predictive performance as the number of features grows. In real-world applications, although the total number of candidate features may be large, it is very likely that only a small proportion are conditionally associated with the outcome. This smaller number, $D$, of \emph{active features} should be the primary driver of the difficulty of the problem, in a minimax sense. Sparse estimation \cite{hastie2015statistical,tibshirani1996regression} is a vast field addressing such data science problems and developing effective estimation procedures, which is especially interesting when the total number of features, $d$, is much larger than $D$.

In this paper, we consider a nonparametric procedure that can simultaneously select important features and estimate the conditional mean function (using only those selected features). For this procedure, the estimation error scales favorably with total dimension (proportional to $log(d)$). Moreover, engaging with a tensor product space additionally means that our active dimension $D$, only shows up multiplicatively in a $log^{D}(n)$ term (as compared to modifying the rate of convergence in $n$ in classical multivariate Sobolev/Holder spaces). Finally, our proposed framework is also seen to be effective in our data example comparisons in Section~\ref{section:numerical}. 

The proposed method considers penalized sieve estimation in multivariate tensor product spaces. Sieve estimation (also known as projection estimation \cite{introtononpara}) is a classical estimation strategy that has been shown to be very effective in univariate regression problems. Sieve estimators can suffer in classical multidimensional spaces (as a large number of basis vectors are required). In this work, we show that in tensor product spaces, some of these issues are alleviated.


\textbf{Notation:} In this paper, we will use bold letters to emphasize a Euclidean vector $\mathbf{x}\in\mathbb{R}^d$ when its dimension $d$ is strictly greater than $1$. The notation $\mathbf{x}^k\in\mathbb{R}$ means the $k$-th entry of $\mathbf{x}\in \mathbb{R}^d$ (rather the $k$-th power of it). We use $\mathbb{N}$ to represent the set of non-negative integers and use $\mathbb{N}^+$ for strictly positive integers. The $\mathbb{N}^d$ is the set of $d$-tuple grids.

\vspace{-3mm}

\section{Univariate Nonparametric Regression and Sieve Estimation}
\label{section:univariatesieve}

One can frame the goal of regression as estimating the function $f$ that minimizes the population mean-squared error (MSE): $E[(Y-f(\mathbf{X}))^2]$, where $Y$ is our outcome, and $\mathbf{X}$ are our features. We denote the distribution of $X$ as $\rho_X$. The minimizer is the well-known condition mean function $f^0(\mathbf{X}) = E[Y | \mathbf{X}]$. In nonparametric regression, we assume $f^0$ belongs to some regular function spaces. An informative univariate model space that we will engage with is the $1^{st}$-order Sobolev space $W_1([0,1])$: 
\begin{equation}
\label{eq:W1}
    f^0 \in W_1([0,1]) = \left\{ f\in L_2([0,1])\ \mid \  f'\text{ exists and }f'\in L_2([0,1])\right\}.
\end{equation}

Here $f'$ can be understood as the weak derivative of $f$. Under the general notation, piece-wise liner functions is a subset of $W_{1}([0,1])$. Without loss of generality, we will assume feature $\mathbf{X}$ belongs to the $d$-dimensional unit cube $[0,1]^d$. Sieve estimation for $f^0$ in the $W_1$ space is built upon the following basic fact: it is possible to express $f^0$ as an (infinite) linear combination of some basis functions $\{\phi_j\}$. Among many possibilities, we choose the following function system as a concrete example:
\begin{equation}
\label{eq:phi}
    \phi_1(x) = 1, \phi_j(x) = \sqrt{2}\cos((j-1)\pi x).
\end{equation}
The aforementioned ``infinite linear combination" can be expressed as: $f^0 = \sum_{j=1}^{\infty} \beta_j^0\phi_j$. Moreover, it is also known that the (generalized) Fourier coefficients $\beta_j^0$ decay at a rate faster than $j^{-1.5}$ for $f^0\in W_1([0,1])$. Therefore, it is plausible to truncate the infinite series to a certain finite level $J_n$: Using only the first ``most important'' $J_n$ basis vectors, one can construct an estimator of $f^0$ with relatively small bias. Formally, a sieve estimator $f_n$ takes the form that $\hat f_n= \sum_{j=1}^{J_n}\hat{\beta}_j\phi_j$ where the coefficients are determined using the available training data $\{(\mathbf{X}_i,Y_i), i = 1,...,n\}$. The coefficients can be determined by solving least-square problems \cite{introtononpara} or using stochastic approximation methods \cite{zhang2021sieve}, both strategies would lead to minimal generalization error (in a minimax-rate sense).

\vspace{-3mm}

\section{Multivariate Nonparametric Models}
\label{section:multivariatemodels}

In most real-world problems, we have more than one feature. In addition it is not always a priori clear which model space to use.

The nonparametric additive model \cite{friedman1981projection} has been seen as one of the most direct models for multivariate nonparametric learning problems. There, we assume features do not interact, or more formally that the regression function take the following additive form:
\begin{equation}
    f^0(\mathbf{x}) = \sum_{k=1}^d f^0_k(\mathbf{x}^k),\quad f^0_k\in W_1([0,1]).
\end{equation}
The lack of interaction between features makes the additive model quite restrictive. There are also some very flexible models widely discussed in the literature, such as Sobolev-type smooth function spaces. Formally, let $\mathbf{a} = (\mathbf{a}^1,...,\mathbf{a}^d)\in (\mathbb{N})^d$, we define the (weak) partial derivative function $D^{\mathbf{a}} f$ of $f$ as:
\begin{equation}
D^{\mathbf{a}} f=\frac{\partial^{\|\mathbf{a}\|_1}}{\partial x_{1}^{\mathbf{a}^1} \cdots \partial x_{d}^{\mathbf{a}^d}} f, \text{ where }\|\mathbf{a}\|_1 = \sum_{k=1}^d \mathbf{a}^k. 
\end{equation}
In this notation, we assume $f^0$ satisfies the following smoothness conditions:
\begin{equation}
\label{eq:isosobolev}
f^0\in W_s\left([0,1]^{d}\right)=\left\{f \in L_2\left([0,1]^{d}\right)\ \mid \  D^{\mathbf{a}} f \in L_2\left([0,1]^{d}\right) \text { for all } \|\mathbf{a}\|_1\leq s\right\}.
\end{equation}

These types of smooth classes do not explicitly assume any specific form such as additivity, but as a cost, suffer substantially more from the ``curse of dimensionality". Although less likely to be miss-specified, this type of model is sometimes thought to be too large to explain the success of many machine learning methods, or be directly applied.

\subsection{Tensor Product Models}
\label{section:tensorproductmodels}

Additive models (mentioned earlier) are an attractive approach for extending univariate smooth functions to multivariate regression. If the true regression function is nearly additive, then with a relatively small number of samples, one can fit a strong additive estimate. However, in some applications it may be that there are important ``interactions'' to consider.  One natural extension to the additive model is to include product-terms of basis functions between individual features. For example, we may consider:
\begin{equation}
\label{eq:additiveexample}
    f^0(\mathbf{x}) = \sum_{k=1}^d f^0_k(\mathbf{x}^k) + a(\mathbf{x}^1)b(\mathbf{x}^2) + c(\mathbf{x}^1)d(\mathbf{x}^3)+e(\mathbf{x}^1)f(\mathbf{x}^2)g(\mathbf{x}^3)+\dots,
\end{equation}
where all the univariate functions above belong to class of smooth functions $W_1([0,1])$. This type of models has been studied in the literature as \emph{Tensor Product Space models} \cite{lin2000tensor}. In more compact notation:
\begin{equation}
\label{eq:S1algebriac}
   f^0 \in  S_1([0,1]^d) = \left\{f = \sum_{m=1}^N \prod_{k=1}^d f_{mk}(\mathbf{x}^k) \text{ for some finite }N\text{, and }f_{mk}\in W_1([0,1])\right\}.
\end{equation} 

Although we defined the $S_1$ space in \eqref{eq:S1algebriac} by addition and multiplication of univariate regular functions, there is an (almost) equivalent characterization of it in the language of partial derivatives:
\begin{equation}
\label{eq:dominatingmixed}
    S_1([0,1]^d) = \left\{f \in L_2\left([0,1]^{d}\right)\ \mid \  D^{\mathbf{a}} f \in L_2\left([0,1]^{d}\right) \text { for all } \|\mathbf{a}\|_{\infty}\leq 1\right\}.
\end{equation}
Function spaces similar to \eqref{eq:dominatingmixed} are called Sobolev spaces with dominating mixed derivatives. They are also characterized as the tensor product spaces of univariate Sobolev spaces $W_1([0,1])$. Compared with the (isotropic) Sobolev spaces defined in \eqref{eq:isosobolev}, tensor product spaces may appear to be formally similar, but have different (and favorable) properties related to statistical estimation. For function space $W_1([0,1]^d)$, we required regular partial derivatives for any index $\mathbf{a}$ satisfying $\|\mathbf{a}\|_1\leq 1$. But for tensor product space $S_1([0,1]^d)$, we require partial derivatives for those indices satisfying $\|\mathbf{a}\|_{\infty} \leq 1$. The latter requirement is strictly stronger and as the dimension $d$ increases, the difference between these two requirements becomes more meaningful. At the same time, the $S_1([0,1]^d)$ space requires less regularity than the $d$-th order isotropic Sobolev space $W_d([0,1]^d)$. In particular, assuming $f^0 \in W_d([0,1]^d)$ means that $\frac{\partial^d}{\partial^d \mathbf{x}^k} f^0$ exists and is square-integrable for any $k = 1,2,...,d$, however functions in $S_1([0,1])$ space do not
need to have second partial derivatives $\frac{\partial^2}{\partial^2 \mathbf{x}^k} f$ for any $k$ (so piece-wise linear functions can be elements of $S_1([0,1]^d)$). More formally, we have the following inclusion relationship:
\begin{equation}
\label{eq:inclusion}
    W_d([0,1]^d) \subsetneq S_1([0,1]^d) \subsetneq W_1([0,1]^d).
\end{equation}
Functions in $S_1$ are allowed to have different degrees of regularity in different ``directions". Specifically, they have almost minimal smoothness in the coordinate axis directions. They are the directions in which people believe most variation should be observed, which is also supported by the success of additive models in practice.

\section{Least-squares Sieve Estimators in Tensor Product Models}
\label{section:sieveintensorproduct}
Sieve estimation leverages the fact that smooth functions can be written as an infinite linear combination of some basis functions $\phi_j$ and the coefficients decay quickly. To construct estimates, we can use a truncated series to balance the approximation and estimation errors. Since functions in $S_1([0,1]^d)$ can be approximately written as the addition and multiplication of a set of univariate functions in $W_1([0,1])$, we may expect a function $f\in S_1([0,1]^d)$ to have the expansion
\begin{equation}
    f(\mathbf{x}) = \sum_{\mathbf{j} \in (\mathbb{N}^+)^d} \beta^0_{\mathbf{j}} \psi_{\mathbf{j}}(\mathbf{x}), \text{ for some }\beta^0_{\mathbf{j}}\in\mathbb{R},
\end{equation}
where $ \mathbf{j} = (\mathbf{j}^1, \mathbf{j}^2,..,\mathbf{j}^d)\in(\mathbb{N}^+)^d$, and $\psi_{\mathbf{j}}$ is a product of the univariate cosine basis $\ \psi_{\mathbf{j}}(\mathbf{x}) = \prod_{k=1}^d \phi_{\mathbf{j}_k}(\mathbf{x}^k)$.

In contrast to the univariate case, there is no single obvious ``natural ordering" of the basis functions $\psi_{\mathbf{j}}$ since they are indexed by some $d$-tuples $\mathbf{j}$. To apply sieve estimation in tensor product spaces (or for any multivariate nonparametric models), we need to establish an order on $\{\psi_{\mathbf{j}}\}$ and determine which basis functions should be used for each $n$. In other words, we need to unravel the \textit{set} $\{\psi_{\mathbf{j}}, \mathbf{j}\in (\mathbb{N}^+)^d\}$ to a \textit{sequence} of functions $\{\psi_j, j \in \mathbb{N}^+\}$. They contain the same set of functions but the latter is an ordered set.

Let $\psi_j$ be the sequence of functions unravelled from $\psi_{\mathbf{j}}$ (we postpone the details of the rearrangement rule to Section~\ref{section:unravelling}). In the new notation, any $f\in S_1([0,1]^d)$ has the expansion $f(\mathbf{x}) = \sum_{j=1}^{\infty} \beta_j^0\psi_j(\mathbf{x})$, $\beta_j^0 \in \mathbb{R}$. To perform sieve estimation in $S_1([0,1]^d)$, we also truncate the series at a proper level $J_n$. The least-square sieve estimator $f_n^{OLS}$ is
    $f_n^{OLS}(\mathbf{x}) = \sum_{j=1}^{J_n} \beta_{jn}^{OLS}\psi_j(\mathbf{x})$
, whose coefficients are the minimizers of the following empirical least-squares problem:
\begin{equation}
\label{eq:olscoeff}
    \left(\beta_{1n}^{OLS},...,\beta_{J_nn}^{OLS}\right) = \operatorname*{argmin}_{(\beta_1,...,\beta_{J_n})\in \mathbb{R}^{J_n}} \sum_{i=1}^n \left(Y_i - \sum_{j=1}^{J_n} \beta_{j}\psi_j(\mathbf{X}_i)\right)^2
\end{equation}

Using standard analysis tools from empirical process theory, it is possible to derive some theoretical guarantee regarding the performance of $f_n^{OLS}$: For $f^0\in S_1([0,1]^d)$,
\begin{equation}
\label{eq:projectionest}
    E\left\|f_n^{OLS} - f^0 \right\|_{L_2(\rho_X)}^2 = O\left(\left(\frac{\log^{d-1}(n)}{n}\right)^{\frac{2}{3}}\log(n)\right),
\end{equation}
when the number of basis function $J_n$ is chosen to be $J_n = \Theta\left(n^{\frac{1}{3}}\log^{\frac{2(d-1)}{3}}(n)\right)$. The proof of the above statement is very similar to that of Theorem~1 in \cite{zhang2021online}, combined with approximation results that are given Lemma~\ref{lemma:approximationinellipsoid}. The above theoretical guarantee is almost minimax-optimal \cite{lin2000tensor}, up to a logarithm term.

The generalization MSE of this least-squares sieve estimator only differs from $n^{-2/3}$ (the rate for $d=1$) by a polylog term (with the dimension $d$ in the exponent). 
This is much improved as compared with estimation in spaces such as $W_s([0,1]^d)$. For those classical spaces, the minimax rate is of the order $n^{-\frac{2s}{2s+d}}$. The dimension $d$ shows up in the exponent of $n$ rather than $\log n$. That horrible dependence on the dimension is one manifestation of the ``curse of dimensionality''. It is much alleviated, as we have shown, in tensor product spaces. Many semiparametric procedures require convergence of intermediate components at a rate of at least $n^{-1/2}$ (e.g., \cite{kennedy2022semiparametric}). Classical Sobolev models must assume $s \geq  d/2$ to give such a guarantee. This requirement may be too strong for many applications: specifically, it already excludes all the piece-wise linear truth when $d \geq 4$. 

\subsection{Important Technical Details: Unravelling}
\label{section:unravelling}
In this section, we are going to talk about how to rearrange a set of functions $\psi_{\mathbf{j}}$ indexed by $d$-tuple to a sequence of functions $\psi_j$. For ease of discussion, we will term this kind of rearrangement process as ``unravelling." Now we present our proposed unravelling rule for tensor product models. 

In Figure~\ref{fig:unravelling}, we present how to rearrange the grids into a sequence. Here we take $d = 2$ as an example. We consider the function $c_{\mathbf{j}} = \mathbf{j}^1\cdot \mathbf{j}^2$ for each grid element $\mathbf{j}\in (\mathbb{N}^+)^2$. We rearrange the grid on the left based on the product, $c_{\mathbf{j}}$, value assigned to them in increasing order. In the right panel of Figure~\ref{fig:unravelling}, we can see grid-elements assigned with smaller $c_{\mathbf{j}}$ values get a more prioritized position in the sequence indexed by $j\in \mathbb{N}^+$. For example, $(1,1)$ is mapped to the first element on the right because it has the smallest product. In contrast, $(2,2)$ gets the $7$-th position. For grids with the same $c_{\mathbf{j}}$ values (such as $(1,2)$ and $(2,1)$), their relative order can be defined arbitrarily. We put $(2,1)$ in front of $(1,2)$ because it has a larger value in the first dimension. In many parts of the theoretical analysis, we are interested in the magnitude of the unravelled sequence $c_j$ (the series presented in right panel of Figure~\ref{fig:unravelling}).

This unraveling in $(\mathbb{N}^+)^d$, gives a rule for arranging the basis functions $\{\psi_{\mathbf{j}}\}$ in a sequence $(\psi_j)$. Using the $c_{\mathbf{j}}$-unravelling rule presented in Figure~\ref{fig:unravelling}, the first several basis functions in the unravelled sequence are $\psi_1 = \psi_{(1,1)}$, $\psi_2 = \psi_{(2,1)}$, $\psi_3 = \psi_{(1,2)}$ and $\psi_7 = \psi_{(2,2)}$. These are the basis functions we used in constructing least-square sieve estimators in \eqref{eq:olscoeff}. Now we present a formal discussion of the above $c_{\mathbf{j}}$-unravelling rule:

\begin{figure}[t!]
    \centering
    \includegraphics[width = 0.6\textwidth]{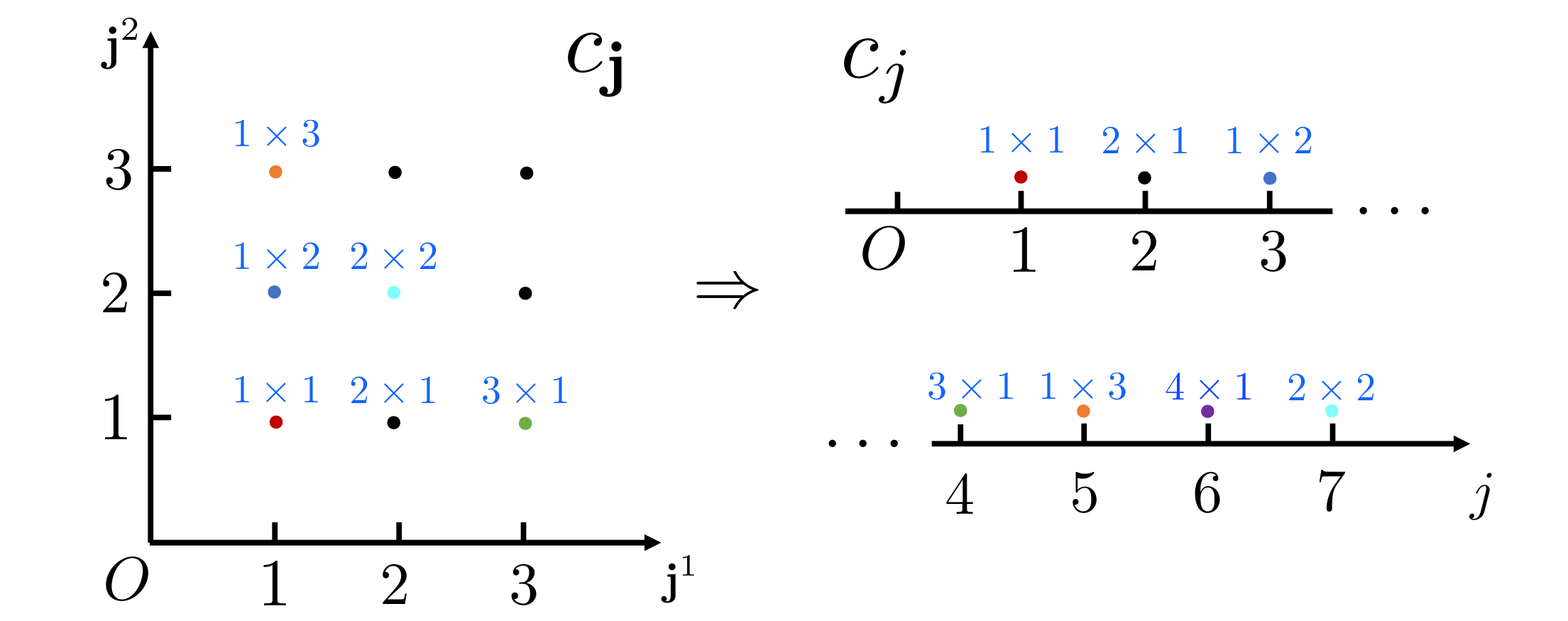}
    \caption{Illustration of unravelling. The rule function is $c_{\mathbf{j}} = \prod_{k=1}^d \mathbf{j}^k$.}
    \label{fig:unravelling}
\end{figure}
\begin{definition}
\label{def:unravelling}
Given a function $c:(\mathbb{N}^+)^d \rightarrow\mathbb{N}^+$ defined on the $d$-tuple grids , we define $\mathcal{U}(\mathbf{m}) :(\mathbb{N}^+)^d\rightarrow\mathbb{N}^+$ to be the unique surjective mapping satisfying the following conditions:
\begin{enumerate}
    \item $\mathcal{U}(\mathbf{m}) \leq \mathcal{U}(\mathbf{n}) \text{ if and only if } c_{\mathbf{m}} \leq c_{\mathbf{n}}$;
    \item (tie-breaker) For $\mathbf{m},\mathbf{n}\in (\mathbb{N}^+)^d$ that have the same $c$ values: $c_{\mathbf{m}} = c_{\mathbf{n}}$, we set $\mathcal{U}(\mathbf{m}) < \mathcal{U}(\mathbf{n})$ if and only the following condition holds: There exists a $k\in \{1,2,...,d\}$ such that,
$
        \mathbf{m}^l = \mathbf{n}^l \text{ for all }l \leq k\text{, but }\mathbf{m}^k > \mathbf{n}^k.
$
\end{enumerate}
We call such a mapping, $\mathcal{U}$, the $c_{\mathbf{j}}$-unravelling rule.
\end{definition}

Condition 1 in Definition~\ref{def:unravelling} is essential: grids with smaller $c_{\mathbf{j}}$ values get a more prioritized position in the unravelled sequence. Condition 2 is an arbitrary tie-breaking rule and could be modified. 

\section{Penalized Sieve Estimators in Sparse Models}
\label{section:sparsesieve}

In this section, we will discuss how to apply $l_1$-penalized sieve estimators for nonparametric sparse models. The difference between this section and the previous is analogous to the difference between sparse additive models \cite{raskutti2012minimax} and additive models (Section~\ref{section:multivariatemodels}), though the technical tools employed differ somewhat. 

Although there may be a substantial number of features collected, it is common that only a small proportion have some significant association with the outcome variable. We will show that, similar to many other ``sparse" methods, our proposed method is relatively robust to the ambient dimension $d$. It is the active dimension of the problem that has a significant impact. We now formalize the nonparametric sparse model to facilitate our discussion:

\begin{itemize}
    \item[(\bf{SpS})] There exists a $D$-variate function $f^*:[0,1]^D\rightarrow \mathbb{R}$ such that: There is set of indices $\{k_1,..,k_D\}\subset\{1,2,...,d\}$ such that for any $\mathbf{u}~\in~ [0,1]^d$: $ f^0(\mathbf{u}) = f^*(\mathbf{u}^{k_1}, \mathbf{u}^{k_2}, ...,\mathbf{u}^{k_D})$. Moreover, we assume $f^*\in S_1([0,1]^D)$.
\end{itemize}

The first part in (SpS) formally states that there are $D$ features that have dominating association with the outcome; The second condition is a smoothness assumption, which can be potentially replaced by other nonparametric models. Here, we take the $S_1$ space as an example for presenting our ideas, for general discussion see (SpS') in Appendix~\ref{app:appinsparsemulti}.

In the last section, we discussed the need to order the basis functions using the unravelling rule $c_{\mathbf{j}} = \prod_{k=1}^d \mathbf{j}^k$. In the sparse model setting, the unravelling rule is very similar except that we allow ourselves to remove some higher-order interaction terms for computational ease. In particular, we begin with a conservative guess $D^{'}$ for the active dimension $D$. We then remove any interactions of order $> D^{'}$. So long as $D \leq D^{'}$ this will not affect the theoretical performance of our estimator. More formally, our new unravelling rule is:

\begin{itemize} 
    \item[(\bf{SpTB})] Let $\{\phi_j\}$ be the univariate cosine basis: $\phi_1(x) = 1$, $\phi_j(x) = \sqrt{2}\cos((j-1)\pi x)$. Consider their natural $d$-dimensional product extension $\psi_{\mathbf{j}}(\mathbf{x}) = \prod_{k=1}^{d}\phi_{\mathbf{j}^k}(\mathbf{x}^k)$, denote $\psi_j$ to be the $c_{\mathbf{j}}$-unravelling sequence of $\{\psi_{\mathbf{j}}\}$. The unravelling rule $c_{\mathbf{j}}$ is defined as
    \vspace{-2mm}
    \begin{equation}
    \label{eq:sparseunravelling}
        c_{\mathbf{j}} = \left\{ \begin{array}{cc}
            \prod_{k=1}^d \mathbf{j}^k &  \text{ , if at most }D' \text{ entries of } \mathbf{j} \text{ are greater than }1 \\
            \infty & \text{, otherwise}
        \end{array}\right.
    \end{equation}
\end{itemize}

Suppose $d = 3$ and we choose the working dimension $D' = 2$. Then $\psi_{(1,1,1)}$ will get the first place when unravelling $\{\psi_{\mathbf{j}}\}$ to the $(\psi_{j})$ sequence. Similarly, $\psi_{(2,1,1)}$ gets the second position and $\psi_{(1,2,1)}$ gets the third. However, basis functions that vary in more than $D'=2$ dimensions will not be used for our estimate. For example, $\psi_{(2,2,2)}(\mathbf{x}) = 2^{3/2} \prod_{k=1}^3 \cos(\pi \mathbf{x}^k)$ is excluded since it varies in all three dimensions. We formalize this using an infinite value of the index in our rule \eqref{eq:sparseunravelling}. 

For problems with higher feature dimension $d$ and limited samples, the empirical least-squares problem \eqref{eq:olscoeff} is likely to be under-determined (we will have more basis functions than samples), and thus regularization is required. In addition, basis functions from non-active features should have $0$ coefficient. Toward this end, we add a sparsity-inducing penalty. More specifically our estimator is given by solving the following penalized optimization problem:
\begin{equation}
\label{eq:LASSOproblem}
\left(\beta_1^{PLS}, ...,\beta_{J_n}^{PLS}\right) = \operatorname*{argmin}_{(\beta_1,..,\beta_{J_n})\in\mathbb{R}^{J_n}} \frac{1}{n}\sum_{i=1}^n \left(Y_i - \sum_{j=1}^{J_n} \beta_j\cdot\psi_j(\textbf{X}_i)\right)^2 + \lambda_n\sum_{j=1}^{J_n}|\beta_j|.
\end{equation}
where our final estimate is given by $f_{n}^{PLS}(\mathbf{x}) = \sum_{j=1}^{J_n} \beta_j^{PLS} \psi_j(\mathbf{x})$. In Appendix~\ref{app:generatingtheindexlist} we include more details on the implementation of the above method. We have the following theoretical guarantee for this estimator's generalization error:
\begin{theorem}
\label{th:maintheorem}
Suppose $\{(\mathbf{X}_i,Y_i)\in [0,1]^d \times \mathbb{R}, i = 1,2,...,n\}$ is an i.i.d. training sample and the true regression function $f^0$ satisfies \textbf{(SpS)}. Let $\epsilon_i = Y_i - f^0(\mathbf{X}_i)$ be sub-Gaussian random noise variables. We further assume that the distribution of $\mathbf{X}$, $\rho_X$, is continuous with a bounded  density function (from above and away from zero), and the working dimension $D'$ in (\textbf{SpTB}) is no smaller than the intrinsic dimension $D$ in (\textbf{SpS}).

Then, for the $l_1$-penalized sieve estimator $f_n^{PLS}$, constructed with basis functions described in (\textbf{SpTB}), we have:
\begin{equation}
\left\|f_n^{PLS} - f^0 \right\|_{L_2(\rho_X)}^2 =     O_p\left(\log(d)\log(n)\left(\frac{\log^{D-1}(n)}{n}\right)^{\frac{2}{3}}\right),
\end{equation}
when $J_n = C(D)d^{D'}n^{1/3} \log^{D'-1} n$ and $\lambda_n = \sqrt{\log(J_n)/n}$.
\end{theorem}

This convergence rate for $f_n^{PLS}$ looks similar to the rate obtained for the unpenalized estimator $f_n^{OLS}$ with two substantial differences: 1) the $\operatorname{log}^{d-1}(n)$ has been replaced by $\operatorname{log}^{D-1}(n)$ which now only involves the active dimension; and 2) The ambient dimension $d$ is only included through a $\operatorname{log}(d)$ term (as is common in sparse regression).

The $l_1$-penalized optimization problem in \eqref{eq:LASSOproblem} can be solved directly using standard lasso solvers such as glmnet \cite{glmnet}. Asymptotically, the time complexity for constructing the above $l_1$-penalized sieve estimator is of order $O(nJ_n) = O(d^{D'}n^{4/3}\log^{D'-1}n)$. In contrast, standard applications of reproducing kernel ridge regression require $\Theta(n^3)$ computation and give no adaptivity guarantees under feature sparsity. Computationally, the proposed sieve estimator is more suitable for large data sets. Other theoretically competitive methods, such as highly adaptive lasso \cite{benkeser2016highly}, require solving optimization problems that scale as $2^dn$, which is substantially more resource intensive than the proposed method.
\vspace{-2mm}

\section{Numerical Examples}

\label{section:numerical}
In this section we demonstrate the finite-sample performance of the proposed methods using simulated and real data sets. The methods discussed in this manuscript, penalized and least-square sieve estimators, are implemented in the R package \texttt{Sieve}. Currently, the package is available on GitHub (terrytianyuzhang/Sieve).

We first present some numerical results based on simulated data sets. In this section we will consider two types of true regression functions: polynomial truth with second order interaction $f^0_{poly}$ and a trigonometric truth with third order interaction $f^0_{cos}$. We give an additional example in Appendix~\ref{app:moresimulation} where the true conditional means only have some ``interaction" terms of the features under consideration. Under this (more artificial) setting the proposed methods perform much better than tree-based methods. The polynomial truth we will consider is defined as:
\vspace{-4mm}
\begin{equation}
    f^0_{poly}(\mathbf{x}) = \sum_{k=1}^{D-1} Leg(2(\mathbf{x}^k-0.5), 3) + Leg(2(\mathbf{x}^k - 0.5), 2)\cdot Leg(2(\mathbf{x}^{k+1} - 0.5), 2)
    \vspace{-4mm}
\end{equation}
where the $Leg(x,j)$ function is the $j$-th Legendre polynomial $Leg(x,2) = x,\quad Leg(x,3) = (3x^2 - 1)/2$. The trigonometric truth is defined as $f^0_{cos}(\mathbf{x}) =  \sum_{\mathbf{j}\in(\mathbb{N}^+)^d, c_{\mathbf{j}} \leq 8} \ \prod_{k=1}^D\cos((\mathbf{j}^k - 1)\pi \mathbf{x}^k)$ with $c_{\mathbf{j}} = \prod_{k=1}^d \mathbf{j}^k$. 

\begin{figure}
    \centering
    \includegraphics[width = .8\textwidth]{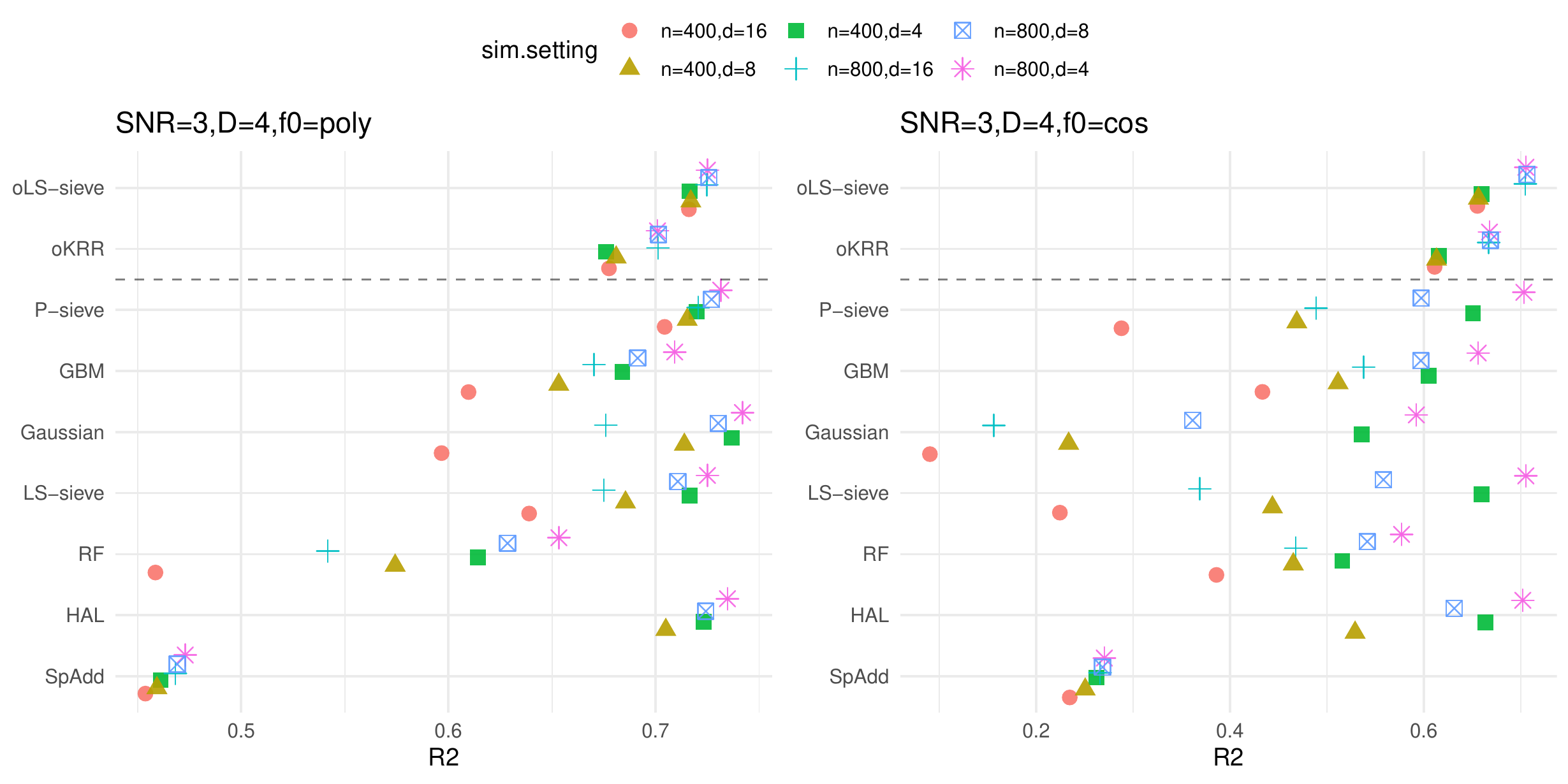}
    \caption{Simulation study results. SNR $ = 3$. Methods above the dashed lines are oracle estimators. }
    \label{fig:SNR3R2}
\end{figure}

The regression estimators we applied in the simulation study are: sieve estimators proposed in this work (least-square and penalized), random forest (RF, R package randomForest), gradient boosting (GBM, R package gbm), Gaussian kernel ridge regression (radial SVM), highly adaptive lasso (HAL, R package hal9001, only applied for the lower dimension case $d = 4$ due to the exponential memory requirement of this method) and sparse additive model (R package SAM). We also include some oracle estimators that know which $D$ dimensions are truly associated with the outcome $Y$ in order to demonstrate the dimension adaptivity of the other methods. The univariate basis $\phi_j$ we used for sieve estimators are: $\phi_1(x) = 1$, $\phi_j(x) = \sin((j+1/2)\pi x)$ (sine basis, for the $f^0_{cos}$ settings) and $\phi_j(x) = \cos((j-1)\pi x)$ (cosine basis, for all the other truth $f^0$). The oracle kernel ridge regression method uses the reproducing kernel of $S_1([0,1]^D)$, see Appendix~\ref{app:productkernels}. We present the simulation study results in Fig~\ref{fig:SNR3R2}. In this section we only consider $D = 4$ and signal-noise-ratio (SNR) $ = 3$. For more simulation results under more settings, see Appendix~\ref{app:moresimulation}. In Fig~\ref{fig:SNR3R2}, model performance is evaluated via generalization-$R^2$. 

\begin{figure}
    \centering
    \includegraphics[width = 0.8\textwidth]{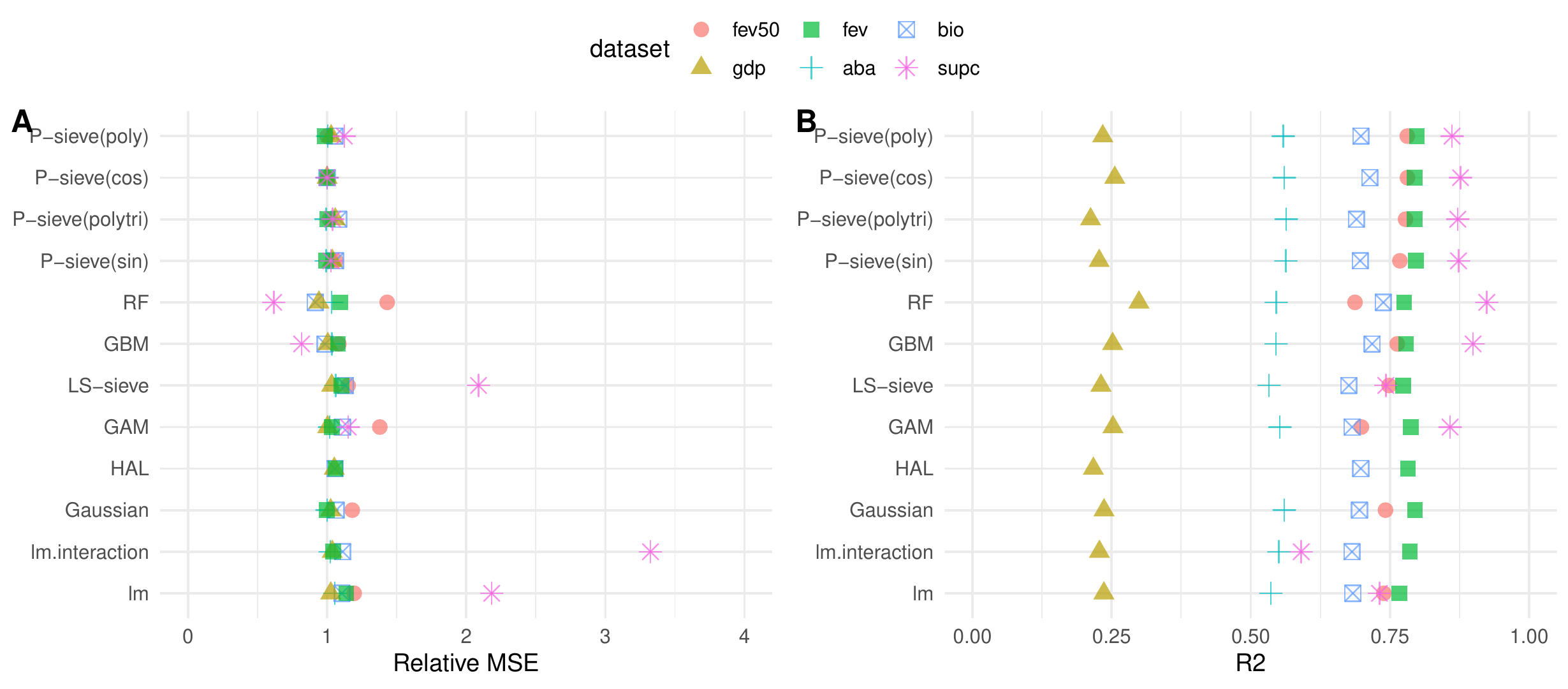}
    \caption{Relative MSE and $R^2$ on real data sets. The MSE values are normalized to that of penalized sieve estimator with cosine functions. Methods requiring significantly more computational resource are not reported.
    }
    \label{fig:realdata}
\end{figure}

We also compare the predictive performance of these methods on 5 public data sets. In Figure~\ref{fig:realdata}, we present the relative testing MSE and (absolute) $R^2$ of each method. We saved $30\%$ of the samples as the test set and the hyperparameters of each method are determined using a 5-fold cross-validation on the training set. The \texttt{fev50} data set uses the true outcome and features of \texttt{fev} together with 50 artificially constructed non-informative features (independent, $\text{Unif}[0,1]$). We use this data set as a sparse feature example. One of the data sets, \texttt{supc}, has been used as an example to demonstrate the effectiveness of tree-based methods such as RF and GBM \cite{hamidieh2018data}, so we also include it for a more comprehensive comparison.

We compared sieve estimators based on different univariate basis $\phi_j$, including Legendre polynomial, cosine and sine basis (the one mentioned earlier in this section), as well as a combination of polynomial and trigonometric functions \cite{Eubank1990CurveFB}. The performance of penalized sieve methods using different basis functions is quite similar. The random forest estimator is more sensitive to the extra dimensions of \texttt{fev50} than penalized sieve and GBM. The proposed penalized sieve method is in general comparable or better than other benchmark methods such as additive models, highly adaptive lasso or Gaussian kernel ridge regression. For more information on the data sets, see Appendix~\ref{app:moresimulation}.

\bibliographystyle{apalike}
\bibliography{main}

\newpage
\appendix
\title{Supplementary Material for\\ Regression in Tensor Product Spaces by the Method of Sieves}
\maketitle

\vspace{-10mm}

\section{More on Numerical Examples and Method Implementation}

\subsection{Supplementary Simulation Details and Results}
\label{app:moresimulation}
In the main text, we present some selected results of our simulation study. In this section we will provide more details. We will also include another truth that only has ``interaction terms". 

In the (complete) simulation study, we considered active dimension $D\in\{2,4\}$ and ambient dimension $d\in\{4,8,16\}$. We used SNR = $3$ and $30$ where the noise random variables $\epsilon$ has a normal distribution. Here SNR is defined as the ratio between the squared 2-norm of $f^0$ and the variance of the noise variables. We choose sample size $n\in\{400,800\}$. The feature vectors $\mathbf{X}$ we consider are uniformly distributed over the $[0,1]^d$ cube. We performed 100 repetitions for each setting. We use oracle hyperparameters for each method of consideration (number of basis functions, regularization parameter, number of trees, etc.), which is determined based on the independent $n = 2000$ testing data set.

\begin{figure}

    \centering
    \includegraphics[width = \textwidth]{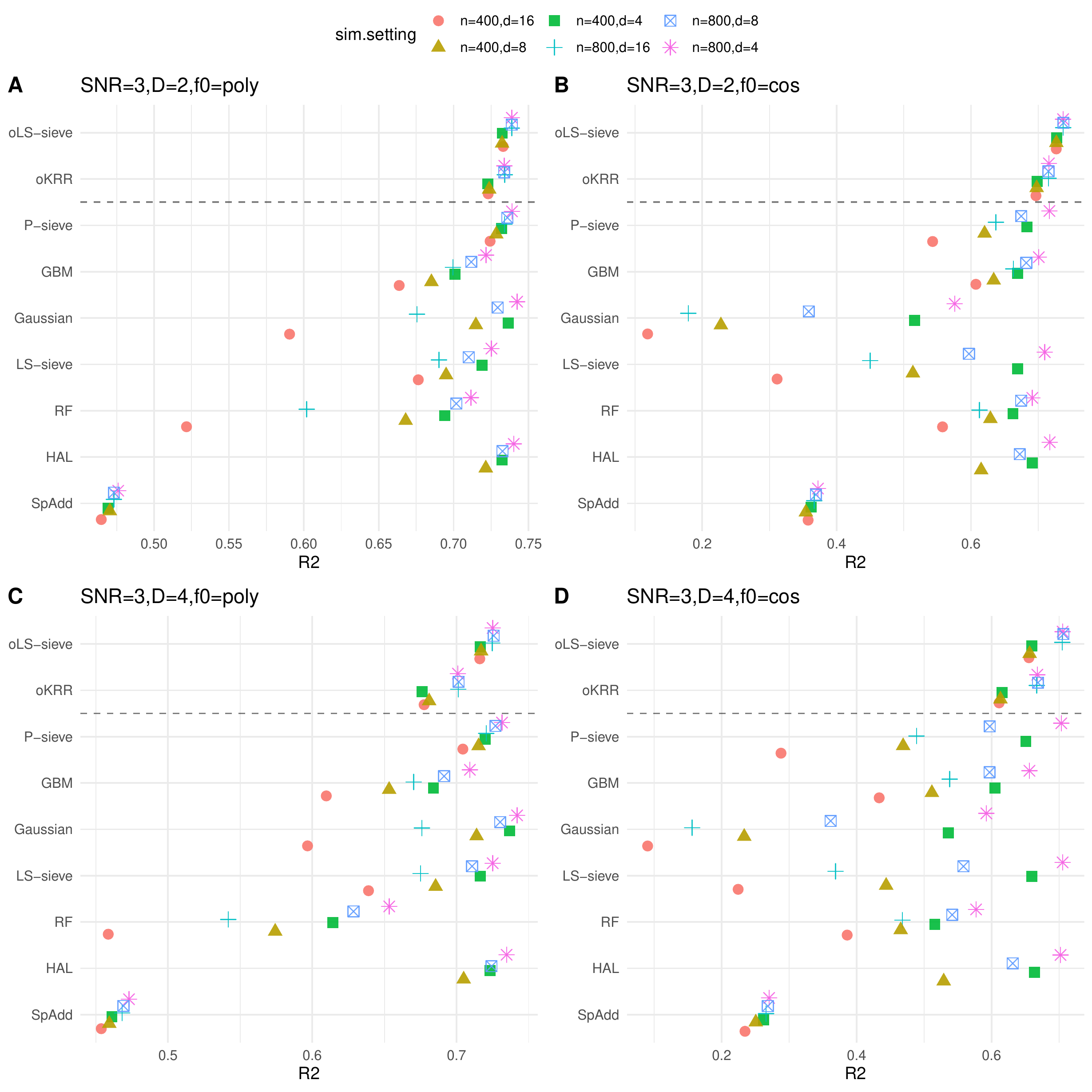}
    \caption{Simulation study results. SNR $ = 3$.}
    \label{fig:SNR3R2supp}
\end{figure}

\begin{figure}
    \centering
    \includegraphics[width = \textwidth]{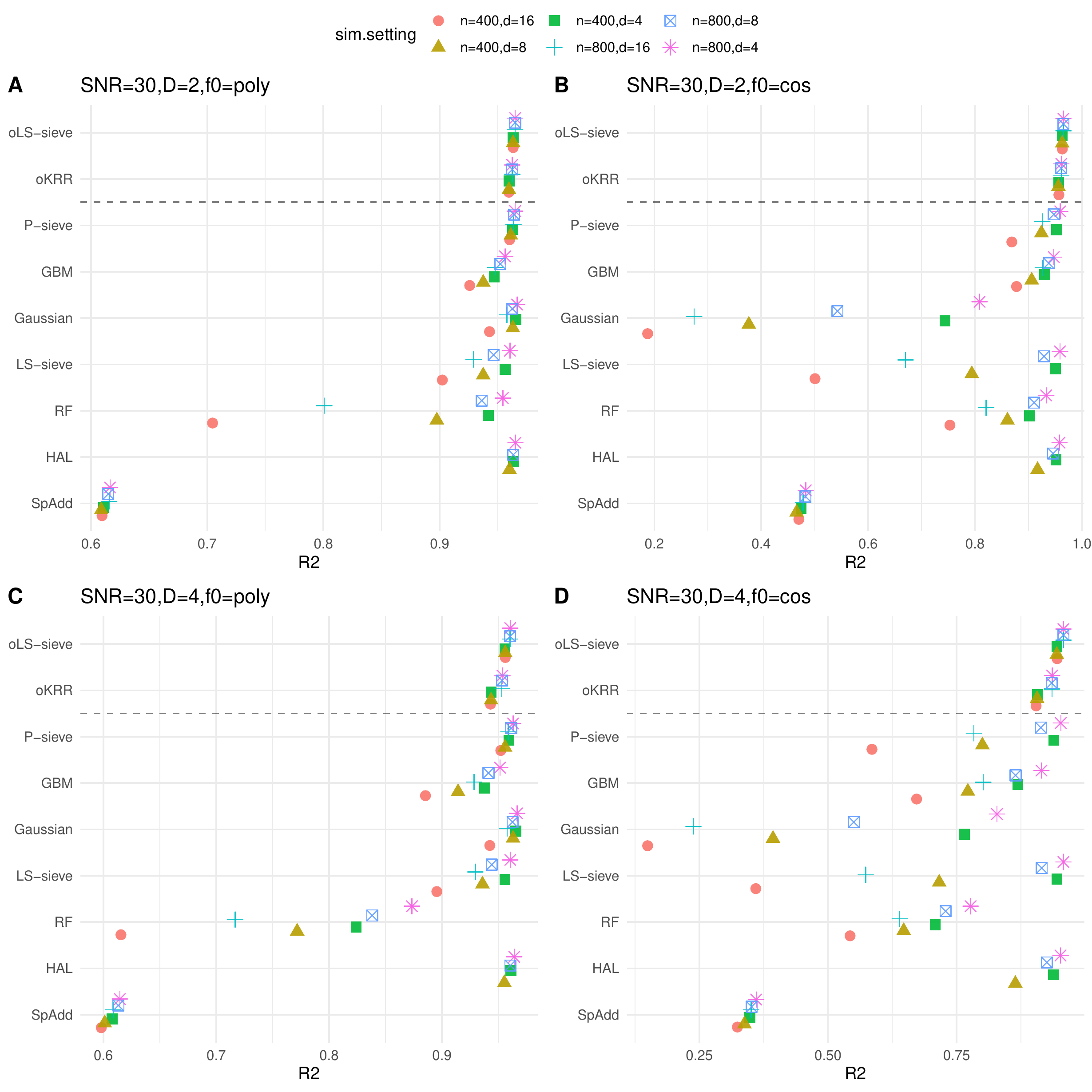}
    \caption{Simulation study results. SNR $ = 30$.}
    \label{fig:SNR30R2supp}
\end{figure}

\begin{figure}
    \centering
    \includegraphics[width = \textwidth]{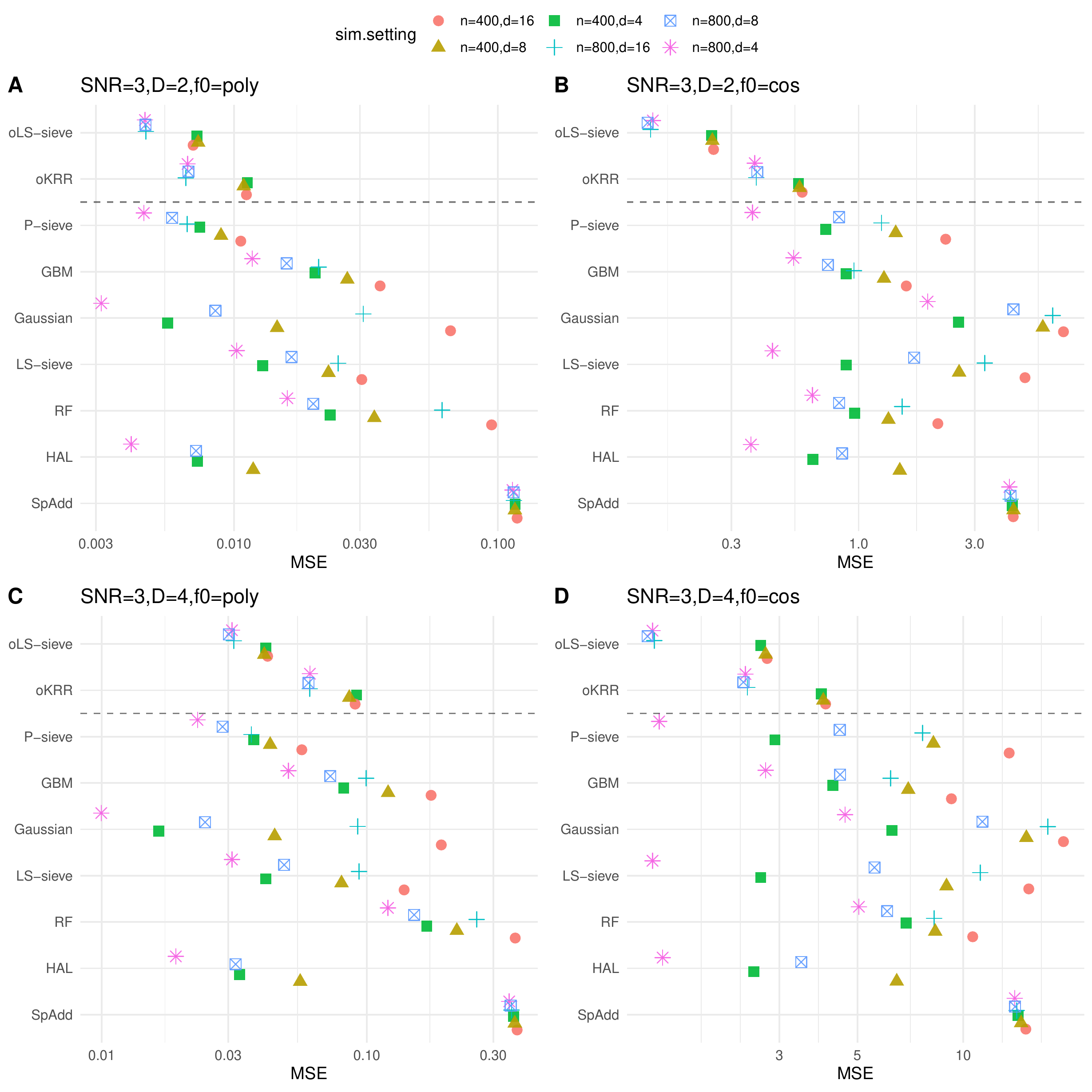}
    \caption{Simulation study results. SNR $ = 3$.}
    \label{fig:SNR3MSEsupp}
\end{figure}

\begin{figure}
    \centering
    \includegraphics[width = \textwidth]{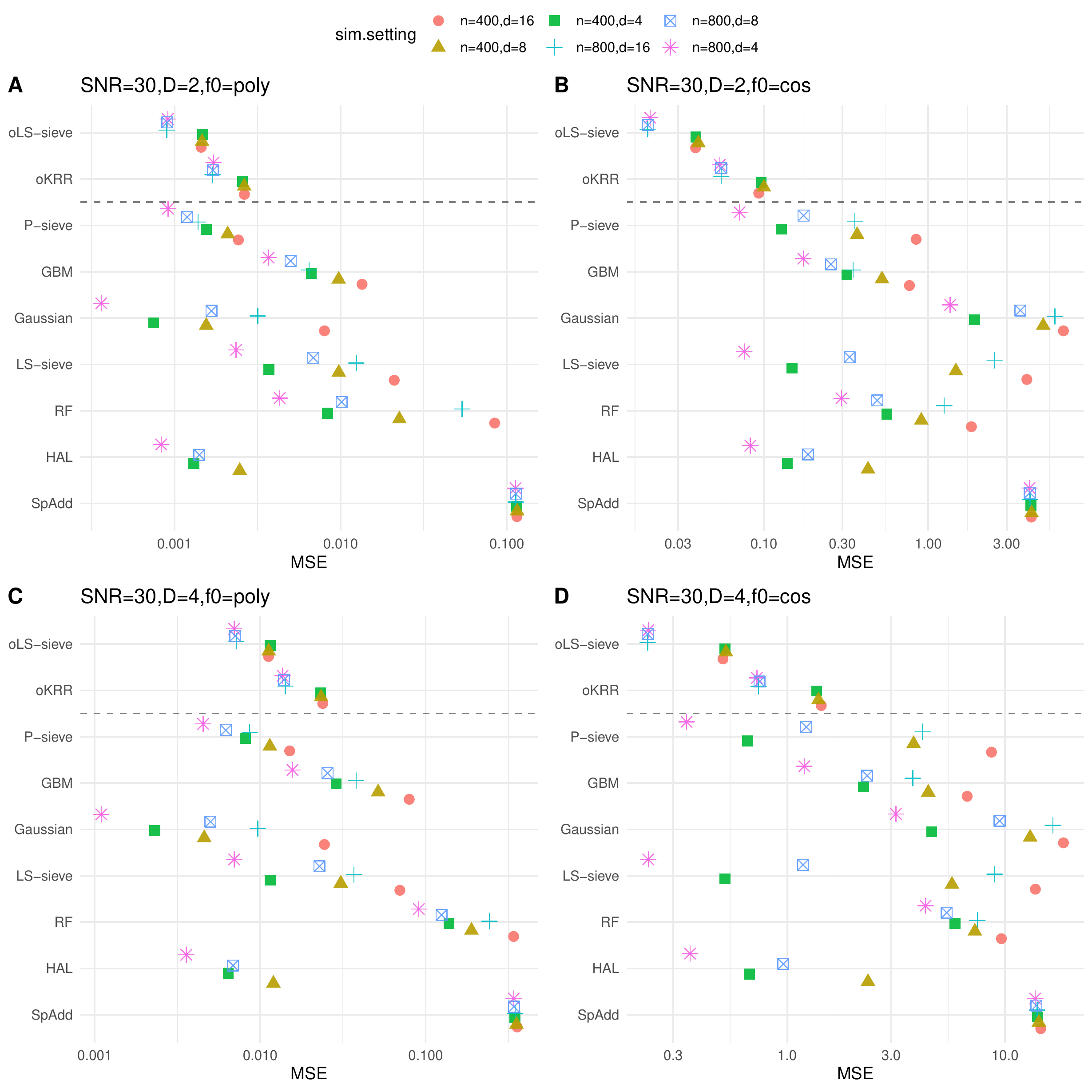}
    \caption{Simulation study results. SNR $ = 30$.}
    \label{fig:SNR30MSEsupp}
\end{figure}

In Figure~\ref{fig:SNR3R2supp}-\ref{fig:SNR3MSEsupp}, we present the simulation results in more settings (as a supplement to the results in Figure~\ref{fig:SNR3R2}). In Figure~\ref{fig:SNR3R2supp} and \ref{fig:SNR30R2supp} the performance is measured in testing $R^2$, in Figure~\ref{fig:SNR3MSEsupp} and \ref{fig:SNR30MSEsupp} it is measured in testing MSE.

\begin{figure}
    \centering
    \includegraphics[width = \textwidth]{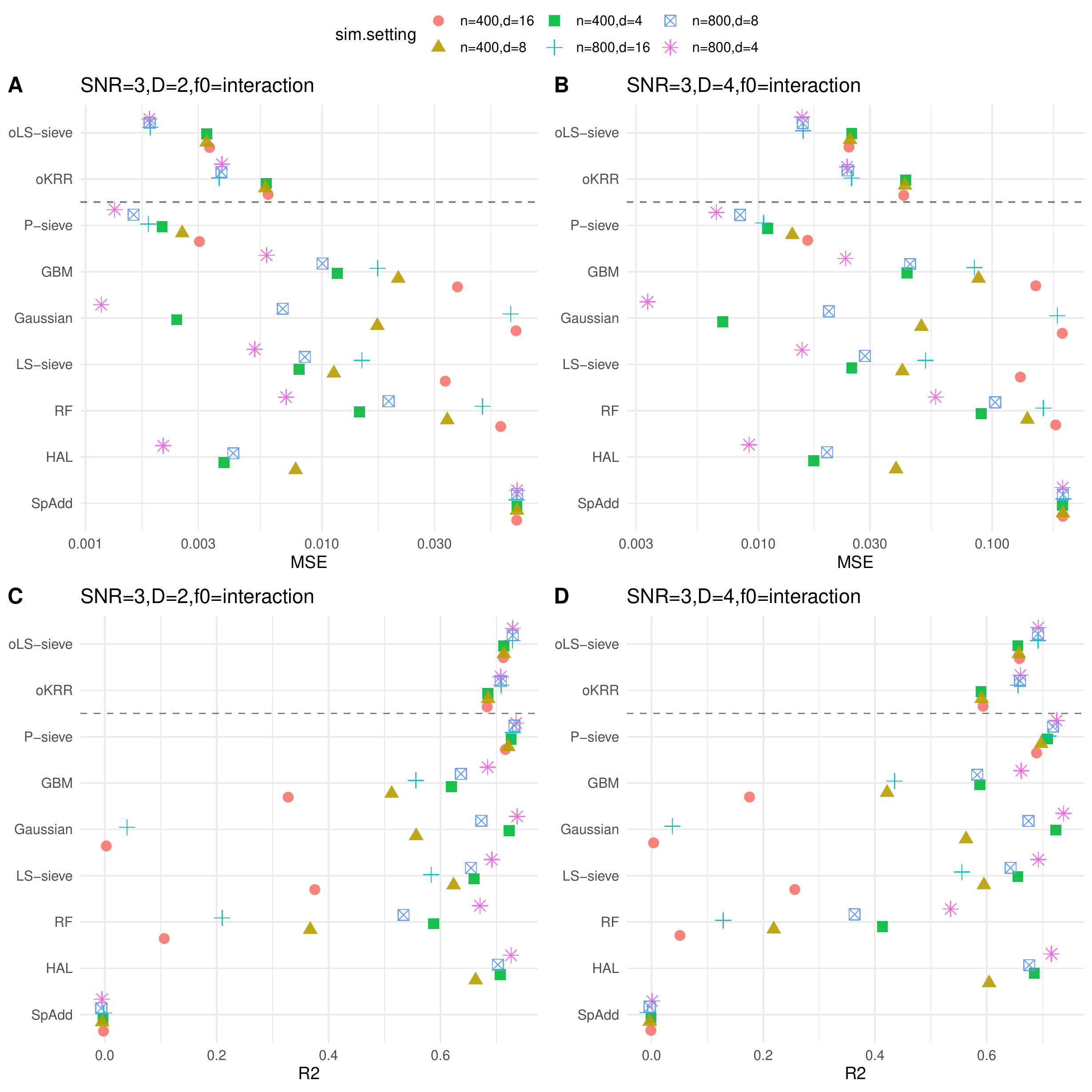}
    \caption{Additional settings, true regression function does not have ``main effect" components. SNR $ = 3$.}
    \label{fig:SNR3interaction}
\end{figure}

\begin{figure}
    \centering
    \includegraphics[width = \textwidth]{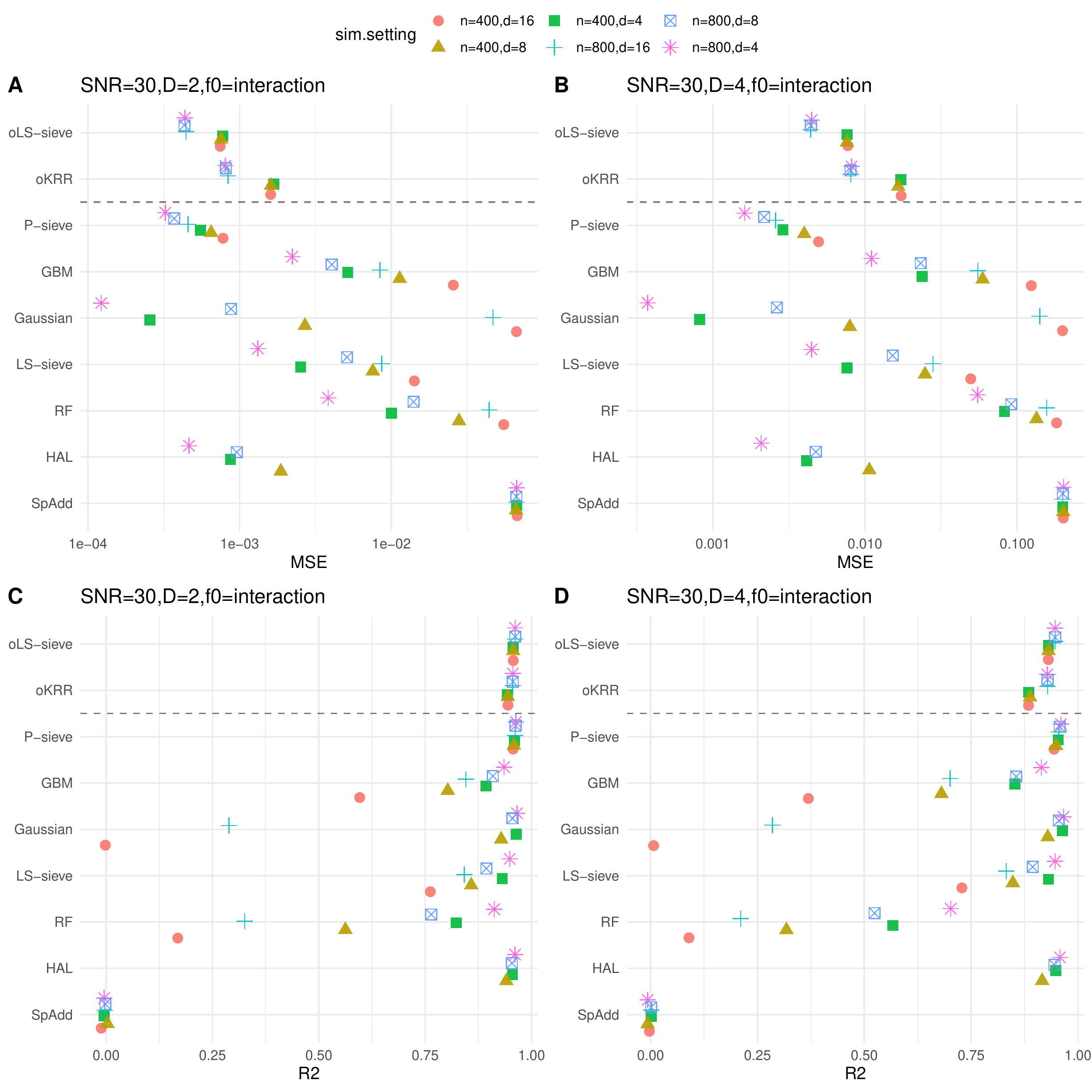}
    \caption{Additional settings, true regression function does not have ``main effect" components. SNR $ = 30$.}
    \label{fig:SNR30interaction}
\end{figure}

We also present the simulation result under a truth that does not have additive component (results are in Figure~\ref{fig:SNR3interaction} and \ref{fig:SNR30interaction}). The truth is:
\begin{equation}
    f^0_{interaction} = \sum_{k=1}^{D-1} Leg(2(\mathbf{x}^k-0.5),2)\cdot Leg(2(\mathbf{x}^{k+1}-0.5),3)
\end{equation}
where the $Leg(x,j)$ function is the $j$-th Legendre polynomial
\begin{equation}
    Leg(x,2) = x,\quad Leg(x,3) = (3x^2 - 1)/2
\end{equation}
This conditional mean is does not depend on any features in a ``main effect" fashion, meaning that $$E[f^0_{interaction}(\mathbf{x})\mid \mathbf{x}^k] = 0$$ for any $1\leq k \leq d$. We can verify this by some direct calculation (recall that $\mathbf{x}\sim Uniform([0,1]^d)$). Although $f^0_{interaction}$ is a simple polynomial with nice smoothness properties, the lack of main effect (or additive component) messes up the performance of many methods. The almost zero testing $R^2$ of additive models demonstrates that in this setting they are no better than taking an unconditional mean of the outcome. Tree-based methods (gradient boosting and random forest) have more difficulties in this setting, especially when compared with their outstanding performance when the main effect components do exist (Figure~\ref{fig:SNR3R2}). Tree-based methods cannot readily decide at which point to divide the feature space. For any binary cut only engaged with one feature, the mean of the outcome on one side of the division should be very similar to that of the other side.

More details of the public real data sets are included in Table~\ref{tb:realdata}. Most of them are available at UCI Machine Learning Repository.

\begin{table}[!t]
\centering
\def~{\hphantom{0}}
\caption{Basic summary of public data sets for performance comparison.}{%
\begin{tabular}{lccc}
 \\
Name & Sample size $n$ & Feature dimension $d$ & Feature type\\
gdp & 616 & 6 & 6 continuous \\
fev & 654 & 4 & 2 continuous, 2 binary\\
fev50 & 654 & 54 & 52 continuous (50 artificial), 2 binary\\
bio & 779 & 9 & 9 continuous\\
aba & 4177 & 8 & 7 continuous, 1 categorical\\
supc & 21263 & 81 & 81 continuous
\end{tabular}}
\label{tb:realdata}
\end{table}

\subsection{Generating the Design Matrices}
\label{app:generatingtheindexlist}
In this section we present more details on efficiently constructing the design matrix for multivariate sieve estimators. In the main text, we discussed the numerical implementation of sieve estimators is reduced to solving a least-square problem or a $l_1$ penalized least-square problem. In both cases we need to construct a design matrix $\hat\Psi$ and store it in the memory. Given a set of multivariate product basis functions $\psi_{\mathbf{j}}(\mathbf{x}) = \prod_{k=1}^{d} \phi_{\mathbf{j}^k}(\mathbf{x}^k) $ indexed by $\mathbf{j}\in(\mathbb{N}^+)^d$, the unravelling rule $c_{\mathbf{j}} = \prod_{k=1}^{d} \mathbf{j}^k$ tells us how to sequentially use them to construct estimators. That is, we have a nonconstructive description of the elements in the unravelled sequence $(\psi_j)$. However, to construct the design matrix $\hat\Psi$ whose elements are $\hat\Psi_{ij} = \psi_j(\mathbf{X}_i)$, we need to know the explicit form of each $\psi_j$. In this section, we aim to create a reference index matrix such that people can easily figure out the analytical form of $\psi_{j}$ by reading through it. In the case $d=D'= 3$, we would have a index matrix $M$ of three columns (corresponding to the three dimensions). The first row has elements: $M_{11} = M_{12} = M_{13} = 1$, corresponding to the constant function $\psi_{(1,1,1)}$. And the following six rows are all $1$ except for $M_{21} = M_{32} = M_{43} = 2$, and $M_{51} = M_{62} = M_{73} = 3$. They corresponds to basis functions $\psi_{(2,1,1)}, \psi_{(1,3,1)}$, etc. By reading through this list, we can directly figure out what are the basis functions needed, and it would facilitate people to construct the design matrices. 

In Algorithm~\ref{algo:GeneratingIndex}, we provide a procedure that can generate such a matrix. By factorizing each natural number as a product of $D'$ numbers sequentially, we can fill out the matrix $M$. In the case when $d = D' = 3$, there is one row having row product equals to $1$, two rows having row product equals to $2$, and six rows having a product equals to $4$. When $D'$ is much smaller than $d$, as for the sparse sieve estimators, there should be much less rows corresponding to the same row product. The algorithm is presented below, followed by an example to explain some of the steps.

\begin{algorithm}[!h]
\label{algo:GeneratingIndex}
\caption{Algorithm for generating the index matrix. For the definition of the $\tau_{D'}$ function mentioned in step *, see Definition~\ref{def:divisorfunction}. In ** step we use the programming language of R to express matrix update.} 
\begin{tabbing}
   \enspace Set maximum row product as $ProdMax$, feature dimension as $d$.\\
   \enspace Define $C^d_m = d!/(m!(d-m)!)$, the combination number of ``choosing $m$ out of $d$ elements".\\
   \enspace $M \leftarrow $ An all $1$ matrix of size $1\times d$.\\
   \enspace FOR $Prod = 2$ TO $Prod = ProdMax$\\
   \qquad Find all the $\tau_{D'}(Prod)$ ways to factorize $Prod$ as a product of $D'$ numbers. *\\
   \qquad Omit all the ``$1$" and combine the same factorizations.\\
   \qquad $GreaterThanOne \leftarrow$ A list. Each element is an array, corresponding to one of the factorizations.\\
   \qquad FOR $i = 1$ TO $i = \text{list length of }GreaterThanOne$\\
   \qquad \qquad $G_i \leftarrow $ The $i$-th element in $GreaterThanOne$.\\
   \qquad \qquad $m\leftarrow \text{The length of the array }G_i$.\\
   \qquad \qquad $Position \leftarrow$ A matrix of size $C(d,m)\times m$.\\
   \qquad \qquad \qquad Each row corresponds to a unique way of choosing $m$ elements from $\{1,2,...,d\}$.\\
   \qquad \qquad $NewIndexMatrix \leftarrow$ A matrix of size $C(d,m)\times d$. All elements are $1$.\\ 
   \qquad \qquad FOR $j = 1$ TO $j = $ $\text{row number of }Position$\\
   \qquad  \qquad  \qquad $NewIndexMatrix[j, Position[j,]] \leftarrow G_i$ **\\
   \qquad  \qquad ENDFOR\\
   \qquad  \qquad $M \leftarrow $ Stack $ M $ above $NewIndexMatrix$ to form a longer matrix.\\
   \qquad  ENDFOR\\
   \enspace  ENDFOR\\
   \enspace RETURN $M$.
\end{tabbing}
\end{algorithm}

We would like to give some examples to better explain the compactly written algorithm above. Let's assume $d = 3$, $D' = 2$. Suppose we are currently at $Prod = 6$ in the first layer of FOR loops. The $\tau_2(6) = 4$ ways to factorize $6$ are:
\begin{equation}
    6 = 6 \times 1 = 1\times 6 = 2\times 3 = 3\times 2.
\end{equation}
After ``Omit all the 1 and combine the same factorizations", we have three ways to factorize $6$ (the first two above are combined to one factorization). Therefore, the $GreaterThanOne$ list looks like
\begin{equation}
    GreaterThanOne = \text{list}([6], [2,3], [3,2]).
\end{equation}
The arrays in $GreaterThanOne$ are of different length. Suppose we are at $i = 2$ in the second layer of FOR loop. Then $G_i = [2,3]$, $m=2$. The $Position$ matrix we constructed is
\begin{equation}
    Position = \begin{bmatrix} 
	1 & 2  \\
	1 & 3 \\
	2 & 3  \\
	\end{bmatrix}.
\end{equation}
This matrix specify at which positions we are going to ``insert" $G_i$. In the inner most FOR loop, we are going to update the all $1$ matrix $NewIndexMatrix$ using the information of $Position$ and $G_i$: $Position$ specifies where to update, $G_i$ specifies to what the elements are updated. When $i = 2, j = 1$, we update the $1^{st}$ and $2^{nd}$ columns in the $1^{st}$ row of $NewIndexMatrix$ to be $[2,3]$, that is
\begin{equation}
    NewIndexMatrix: \begin{bmatrix} 
	1 & 1 & 1  \\
	1 & 1 & 1  \\
	1 & 1 & 1  \\
	\end{bmatrix} \stackrel{\text{Update}}{\longrightarrow}
	\begin{bmatrix} 
	2 & 3 & 1  \\
	1 & 1 & 1  \\
	1 & 1 & 1  \\
	\end{bmatrix}
\end{equation}
When $i = 2, j = 2$, we update the $1^{st}$ and $3^{rd}$ columns in the $2^{nd}$ row of $NewIndexMatrix$ to be $[2,3]$:
\begin{equation}
    NewIndexMatrix: 
    \begin{bmatrix} 
	2 & 3 & 1  \\
	1 & 1 & 1  \\
	1 & 1 & 1  \\
	\end{bmatrix}
	\stackrel{\text{Update}}{\longrightarrow}
	\begin{bmatrix} 
	2 & 3 & 1  \\
	2 & 1 & 3  \\
	1 & 1 & 1  \\
	\end{bmatrix}
\end{equation}
The first several rows of the final $M$ are:
\begin{equation}
	\text{row 1 to 10:}\begin{bmatrix} 
	1 & 1 & 1 \\
	2 & 1 & 1\\
	1 & 2 & 1\\
	1 & 1 & 2\\
	3 & 1 & 1\\
	1 & 3 & 1\\
	1 & 1 & 3\\
	4 & 1 & 1  \\
	1 & 4 & 1  \\
	1 & 1 & 4  \\
	\end{bmatrix}
	\qquad
	\text{row 11 to 20:}\begin{bmatrix} 
	2 & 2 & 1 \\
	2 & 1 & 2\\
	1 & 2 & 2\\
	5 & 1 & 2\\
	1 & 5 & 1\\
	1 & 1 & 5\\
	6 & 1 & 1\\
	1 & 6 & 1  \\
	1 & 1 & 6  \\
	2 & 3 & 1  \\
	\end{bmatrix}
	\qquad
	\text{row 21 to 30:}\begin{bmatrix} 
	2 & 1 & 3 \\
	1 & 2 & 3\\
	3 & 2 & 1\\
	3 & 1 & 2  \\
	1 & 3 & 2  \\
	7 & 1 & 1\\
	1 & 7 & 1  \\
	1 & 1 & 7  \\
	8 & 1 & 1\\
	1 & 8 & 1  \\
	\end{bmatrix}
\end{equation}
So we can read $\psi_1 = psi_{(1,1,1)}$, $\psi_{21} = \psi_{(2,1,3)}$, etc.

\newpage

\section{Product Kernels and Tensor Product Spaces}
\label{app:productkernels}
In this section, we will review the concept of Mercer kernel and reproducing kernel Hilbert space (RKHS). We will first engage with univariate RKHSs and their Sobolev ellipsoid representation in Appendix~\ref{app:UnivariateRKHS}. By considering the tensor product kernel, we can extend our discussion to multivariate tensor product models (Appendix~\ref{app:MultivariateRKHS}). Later in this section, we will arrive at some multivariate Sobolev ellipsoid models. These ``ellipsoids" can be seen as abstractions of the example function spaces (such as $S_1([0,1]^d)$) discussed in the main text. The main purpose of this section is preparing our readers to better understand our motivation of studying multivariate (product) Sobolev ellipsoids in the following sections.

\subsection{Univariate RKHS and Sobolev Ellipsoids}
\label{app:UnivariateRKHS}
There is a vast literature on univariate nonparametric regression problem. Listing a few of them: Sobolev space and spline estimator \cite{wahba1990spline}; reproducing kernel Hilbert space and kernel ridge regression estimator \cite{steinwart2008support}; Sobolev ellipsoid and sieve-type projection estimator \cite{introtononpara}. These function spaces are closely relate to each other: Sobolev space can sometimes be treat as a special case of RKHS and there is always a equivalence between a ball in RKHS and a Sobolev ellipsoid. We will try to give a brief review of this part of nonparametric learning through some examples.

First we are going to present the concept of Mercer-kernels and their related reproducing kernel Hilbert spaces (on real line).

\begin{definition}
A symmetric bivariate function $k:\mathbb{R}\times \mathbb{R}\rightarrow \mathbb{R}$ is positive semi-definite (PSD) if for any $n\geq 1$ and $(x_i)_{i=1}^n\subset \mathbb{R}$, the $n \times n$ matrix $\mathbb{K}$ whose elements are $\mathbb{K}_{ij}= k(x_i,x_j)$ is always a PSD matrix.

A continuous, bounded, PSD kernel function $k$ is called a \emph{Mercer kernel}.
\end{definition}

The following theorem (e.g., from \cite{cucker2002mathematical}) states the existence and uniqueness of a reproducing Hilbert space with respect to a Mercer kernel.

\begin{theorem}
\label{RKHSdefines}
For a Mercer Kernel $k:\mathbb{R}\times \mathbb{R}\rightarrow\mathbb{R}$, there exists an unique Hilbert Space $(\mathcal{H}_k,\langle \cdot,\cdot \rangle_{k})$ of functions on $\mathbb{R}$ satisfying the following conditions. Let $k_x:z\mapsto k(x,z)$:
\begin{enumerate}
    \item For all $x\in \mathbb{R}$, $k_x\in \mathcal{H}_k$.
    \item The linear span of $\{k_x\ \mid \ x\in  \mathbb{R}\}$ is dense (w.r.t $\|\cdot\|_{k}$) in $\mathcal{H}_k$.
    \item For all $f\in \mathcal{H}_k,x\in \mathbb{R}$, $f(x) = \langle f,k_x\rangle_{k}$ (reproducing property).
\end{enumerate}
We call this Hilbert space the \emph{Reproducing kernel Hilbert space (RKHS)} associated with kernel $k$.
\end{theorem}

\begin{example}
The space $W_1([0,1])$ is a RKHS with kernel
\begin{equation}
    k(s,t) = \frac{\cosh(\min(s,t))\cosh(1-\max(s,t))}{\sinh(1)}
\end{equation}
For proof, see \cite{fasshauer2015kernel} (Appendix~A) or \cite{akgul2020new}. And the inner product is defined as
\begin{equation}
   \langle f, g\rangle_{W_1([0,1])} = \int_0^1 f(\tau)g(\tau) d\tau + \int_0^1 f'(\tau)g'(\tau)d\tau
\end{equation}
The reproducing property reads as: for any $x\in [0,1]$ and any $f\in W_1([0,1])$
\begin{equation}
\begin{aligned}
f(x) & =  \langle f, k_x \rangle_{W_1([0,1])} \\
& = \int_0^1 f(\tau)k(x,\tau)d\tau + \int_0^1 f'(\tau)\frac{\partial}{\partial \tau}k(x, \tau)d\tau.
\end{aligned}
\end{equation}
\end{example}

Under mild conditions \cite{steinwart2012mercer}, a Mercer kernel has the following Mercer expansion.
\begin{equation}
\label{eq:mercerexpansion}
  k(s,t) = \sum_{j\in \mathcal{J}}\lambda_j \phi_j(s)\phi_j(t),
\end{equation}
where $\mathcal{J}$ is an at most countably infinite index set. The ``eigenvalues" $\lambda_j$ are real numbers. The ``eigenfunctions" (basis functions) $\{\phi_j\}$ can also be a complete basis of some $L^2$ spaces and the RKHS.

Although the majority of estimation procedures of estimation in RKHS leverages the reproducing properties, the method considered in this paper uses the ``feature maps" directly (which is of a ``sieve" nature). There has been studies shown that considering the problem from this perspective would give much computational advantage over kernel methods \cite{zhang2021online, zhang2021sieve}. Here we present the fundamental connection between a RKHS and a Sobolev ellipsoid established in the literature (e.g., p.37,Theorem 4 in \cite{cucker2002mathematical}).

\begin{theorem}
\label{theorem:equivalentdef}
Under mild conditions, the Hilbert space $\mathcal{H}_k$ of the kernel $k$ (defined in Theorem~\ref{RKHSdefines}) is identical -- same function class with the same inner product -- to the following Hilbert space $\mathbb{H}_k$.
\begin{equation}
    \mathbb{H}_k=\left\{f \mid f=\sum_{j=1}^{\infty} a_{j} \phi_{j} \quad \text { with } \quad \sum_{j=1}^{\infty}  a_j^2 \lambda_j^{-1} < \infty\right\}
\end{equation}
Equipped with the inner product:
\begin{equation}
    \langle f, g\rangle_{k}=\sum_{j=1}^{\infty} \lambda_j^{-1} a_j b_j
\end{equation}
for $f=\sum_j a_{j} \phi_{j}, \text{ and }g=\sum_j b_{j} \phi_{j}$. The functions $\phi_j$ and real numbers $\lambda_j$ are the eigen-system in the Mercer expansion \eqref{eq:mercerexpansion} (assuming $\mathcal{J} = \mathbb{N}^+$).
\end{theorem}

\begin{example}
The reproducing kernel for $W_1([0,1])$ has the following Mercer expansion:
\begin{equation}
    k(s,t) = \sum_{j=1}^{\infty} \phi_j(s)\phi_j(t),
\end{equation}
with
\begin{equation}
\label{eq:cosinebasisapp}
    \begin{aligned}
        \lambda_1 = 1,\quad &\phi_1(x) = 1\\
        ,\quad 
        \lambda_j = \frac{1}{1+((n-1)\pi)^2} & ,\quad \phi_j(x) = \sqrt{2}\cos((n-1)\pi x) \text{ for }j\geq 2
    \end{aligned}
\end{equation}
Therefore, we also have the following characterization of a ``ball" in $W_1([0,1])$:
\begin{equation}
    \{f\in W_1([0,1])\mid \|f\|_{W_1}^2 \leq Q^2\} = \left\{f=\sum_{j=1}^{\infty} a_{j} \phi_{j} \quad \text { with } \quad \sum_{j=1}^{\infty}  a_j^2 \lambda_j^{-1} \leq Q^2 \right\}
\end{equation}
Put in words, a ball in a RKHS is a Sobolev ellipsoid.
\end{example}

\subsection{Multivariate RKHS and Sobolev Ellipsoids}
\label{app:MultivariateRKHS}
Given a univariate RKHS, one of the most naturally related multivariate RKHS is the one corresponding to the ``product kernel". This is also the most common practice of applying multivariate RKHS methods to real-world data sets (noting that multivariate Gaussian kernel is a product of univariate Gaussian kernels). 

\begin{definition}
Given a univariate Mercer kernel $k:\mathbb{R}\times \mathbb{R} \rightarrow \mathbb{R}$, we define its (natural, $d$-dimensional) product kernel $k^d:\mathbb{R}^d\times \mathbb{R}^d \rightarrow \mathbb{R}$ to be:
\begin{equation}
    k^d(\mathbf{s},\mathbf{t}) = \prod_{k=1}^d k(\mathbf{s}^k, \mathbf{t}^k).
\end{equation}
We can also define the RKHS of $k^d$ using the fact that $k^d$ is also a Mercer kernel (Proposition 12.31 of \cite{wainwright2019high}). Typical elements in this multivariate RKHS take the following form:
\begin{equation}
    f(\mathbf{x}) =  \sum_{l=1}^m\prod_{k=1}^d f_{kl}(\mathbf{x}^k)\text{, with }f_{kl}\text{ belong to the univariate RKHS}
\end{equation}
\end{definition}

There are multiple ways to engage with an element in $\mathcal{H}_{k^d}$ and its inner product. One way, as presented above, is using the property that $\mathcal{H}_{k^d}$ is a tensor product Hilbert space of $d$ univariate ones, this would lead to the following characterization of its inner product.

\begin{proposition}
\label{prop:1stinnerproduct}
The RKHS for $k^d$, $\mathcal{H}_{k^d}$, is equipped with the inner product:
\begin{equation}
\label{eq:multiinnerproduct}
\begin{aligned}
     <h, g>_{k^d} = \sum_{k=1}^n \sum_{l=1}^m \prod_{j=1}^d \langle h_{kj}, g_{lj} \rangle_k
\end{aligned}
\end{equation}
for $h(\mathbf{x}) = \sum_{k=1}^n\prod_{j=1}^d h_{kj}(\mathbf{x}^j)$, $g(\mathbf{x}) = \sum_{l=1}^m\prod_{j=1}^d g_{lj}(\mathbf{x}^j)$. The component functions $h_{kj}$, $g_{lj}$ all belong to the univariate RKHS $\mathcal{H}_k$.
\end{proposition}

Alternatively, we can also consider the basis expansion form of the functions in $\mathcal{H}_{k^d}$ (which is very similar to Theorem~\ref{theorem:equivalentdef}). This would lead to some representation related to Sobolev ellipsoid. We first note that the tensor product kernel $k^d$ has the following Mercer expansion (which can be formally verified by direct calculation):
\begin{equation}
    k^d(\mathbf{s},\mathbf{t}) = \sum_{\mathbf{j} \in (\mathbb{N}^+)^d} \prod_{k=1}^d \lambda_{\mathbf{j}^k} \psi_{\mathbf{j}}(\mathbf{s})\psi_{\mathbf{j}}(\mathbf{t})
\end{equation}

We have the following equivalent charascterization:

\begin{proposition}
\label{prop:innerproduct2}
The inner product present in Proposition~\ref{prop:1stinnerproduct} is equivalent to the following one expressed in basis expansion form:
\begin{equation}
    \langle h,g\rangle_{k^d} = \sum_{\mathbf{j}\in(\mathbb{N}^+)^d} \left( \prod_{k=1}^d \lambda_{\mathbf{j}^k}\right)^{-1} h_{\mathbf{j}} g_{\mathbf{j}}
\end{equation}
for $h,g$ in the multivariate RKHS $\mathcal{H}_{k^d}$ with the basis expansion $h = \sum_{\mathbf{j}\in(\mathbb{N}^+)^d} h_{\mathbf{j}} \psi_{\mathbf{j}}$, $g = \sum_{\mathbf{j}\in(\mathbb{N}^+)^d} g_{\mathbf{j}} \psi_{\mathbf{j}}$. The multivariate basis $\psi_{\mathbf{j}} = \prod_{k=1}^d \phi_{\mathbf{j}^k}$ is the product of the eigenfunctions of the univariate kernel $k$ (defined in \eqref{eq:mercerexpansion}).
\end{proposition}

\begin{example}
\label{example:mainexample}
The natural $d$-dimensional tensor product extension of $W_1([0,1])$ space is the RKHS of the kernel:
\begin{equation}
\begin{aligned}
        k^d(\mathbf{s},\mathbf{t}) &= \prod_{m=1}^d k(\mathbf{s}^m, \mathbf{t}^m) = \prod_{m=1}^d k(\mathbf{s}^m, \mathbf{t}^m)\\
    &= (\sinh(1))^{-d}\prod_{m=1}^d \cosh(\min(\mathbf{s}^m,\mathbf{t}^m))\cosh(1-\max(\mathbf{s}^m,\mathbf{t}^m))
\end{aligned}
\end{equation}
The inner product, according to Proposition~\ref{prop:1stinnerproduct}, can be explicitly written as:
\begin{equation}
\begin{aligned}
  <h, g>_{k^d} &= \sum_{k=1}^n \sum_{l=1}^m \prod_{j=1}^d \langle h_{kj}, g_{lj} \rangle_{W_1([0,1])}\\ 
  & = \sum_{k=1}^n \sum_{l=1}^m \prod_{j=1}^d \left(\int_0^1 h_{kj}(\tau)g_{lj}(\tau)d\tau + \int_0^1 h'_{kj}(\tau)g'_{lj}(\tau)d\tau \right)   
\end{aligned}
\end{equation}
for $h(\mathbf{x}) = \sum_{k=1}^n\prod_{j=1}^d h_{kj}(\mathbf{x}^j)$, $g(\mathbf{x}) = \sum_{l=1}^m\prod_{j=1}^d g_{lj}(\mathbf{x}^j)$. The component functions $h_{kj}$, $g_{lj}$ all belong to $W_1([0,1])$. Then the RKHS-norm (induced by the inner product) for a function $h\in \mathcal{H}_{k^d}$ is:
\begin{equation}
\begin{aligned}
        \|h\|_{k^d} 
        & = \sum_{k=1}^n \sum_{l=1}^n \prod_{j=1}^d \left(\int_0^1 h_{kj}(\tau)h_{lj}(\tau)d\tau + \int_0^1 h'_{kj}(\tau)h'_{lj}(\tau)d\tau \right)\\
        & \stackrel{(1)}{=} \sum_{\|\mathbf{a}\|_{\infty} \leq 1} \|D^{\mathbf{a}} h\|_{L_2([0,1]^d)}^2. 
\end{aligned}
\end{equation}
The above step (1) can be checked directly (and using Fubini's theorem). We present the calculation for a simple case where $h(\mathbf{x}) = \prod_{j=1}^d h_{j}(\mathbf{x}^j)$ and $d = 2$:
\begin{equation}
\begin{aligned}
        \|h\|_{k^2} & = \left(\int_0^1 h_1^2(\tau_1)d\tau_1\right)\left(\int_0^1 h_2^2(\tau_2)d\tau_2\right)+ \left(\int_0^1 (h_1(\tau_1))^2d\tau_1\right)\left(\int_0^1 (h_2'(\tau_2))^2d\tau_2\right) +\\
        & \left(\int_0^1 (h_1'(\tau_1))^2d\tau_1\right)\left(\int_0^1 h_2^2(\tau_2)d\tau_2\right) + \left(\int_0^1 (h_1'(\tau_1))^2d\tau_1\right)\left(\int_0^1 (h_2'(\tau_2))^2d\tau_2\right)\\
        & = \int_{[0,1]^2} h^2(\tau_1, \tau_2)d\tau_1 d\tau_2 + \int_{[0,1]^2} \left(\frac{\partial}{\partial \tau_2}h(\tau_1, \tau_2)\right)^2d\tau_1 d\tau_2 + \\
       & \int_{[0,1]^2} \left(\frac{\partial}{\partial \tau_1}h(\tau_1, \tau_2)\right)^2d\tau_1 d\tau_2 + \int_{[0,1]^2} \left(\frac{\partial^2}{\partial \tau_1\partial \tau_2}h(\tau_1, \tau_2)\right)^2d\tau_1 d\tau_2
\end{aligned}
\end{equation}
In a word, $W_1([0,1])$ space is an example of a univariate RKHS; $S_1([0,1])$ space, when equipped with a proper inner product, is the tensor product extension of $W_1([0,1])$. Moreover, Proposition~\ref{prop:innerproduct2} implies an equivalent way to express the RKHS inner product and its induced norm. Specifically, we know that
\begin{equation}
    \sum_{\|\mathbf{a}\|_{\infty} \leq 1} \|D^{\mathbf{a}} h\|_{L_2([0,1]^d)}^2 = \|h\|_{k^d} = \sum_{\mathbf{j}\in (\mathbb{N}^+)^d} \left(\prod_{k=1}^d \mathbf{j}^k\right)^2\beta_{\mathbf{j}}^2 
\end{equation}
for $h = \sum_{\mathbf{j}\in(\mathbb{N}^+)^d} h_{\mathbf{j}} \psi_{\mathbf{j}}$. The multivariate basis $\psi_{\mathbf{j}} = \prod_{k=1}^d \phi_{\mathbf{j}^k}$ is the product of the cosine functions (defined in \eqref{eq:cosinebasisapp}).
\end{example}

In the rest of the paper, we will switch from concrete example spaces to more abstract Sobolev ellipsoid type spaces. The (univariate) Sobolev ellipsoid has been a benchmark model in the literature of sieve estimators, and we just showed how it can be related to some more interpretable spaces. In the multivariate case, we will be engaging with truth function $f^0$ belong to the multivariate Sobolev ``ellipsoid":
\begin{equation}
\label{eq:firstpresentmultisobolev}
    f^0 \in \left\{f = \sum_{\mathbf{j}\in (\mathbb{N}^+)^d} \beta_{\mathbf{j}}\psi_{\mathbf{j}}\ \mid\ \sum_{\mathbf{j}\in (\mathbb{N}^+)^d} \left(\prod_{k=1}^d \mathbf{j}^k\right)^{2s}\beta_{\mathbf{j}}^2 \leq Q^2\right\}.
\end{equation}
for some product basis $\psi_{\mathbf{j}}$. We assumed the regression function can be expanded as an infinite linear combination of a set of basis functions $\psi_{\mathbf{j}}$ indexed by $d$-tuples. And at the same time we require $\beta_{\mathbf{j}}$ to converge to zero at a fast enough rate as the product of index $\mathbf{j}$ goes to infinity. The function space in \eqref{eq:firstpresentmultisobolev} is the same as a ball in some multivariate RKHS (as illustrated in Example~\ref{example:mainexample}). We also introduced another parameter $s$ that determines the decay rate of $\beta_{\mathbf{j}}$, which is often interpreted as a smoothness parameter (\cite{wahba1990spline}, Chapter~2).

\newpage

\section{Unravelling and Approximation Results}
In this section we will first quantify the asymptotic behavior of unravelled series $c_j$ depicted in the right panel of Figure~\ref{fig:unravelling}. We will use these results to reduce Sobolev ellipsoids indexed by a $D$-tuples (such as the one in \eqref{eq:firstpresentmultisobolev}) to ones indexed by a sequence of natural numbers. This will directly lead to some useful approximation results in multivariate function spaces. 
\subsection{Magnitude of Unravelled Series}


In general it is very hard to have an analytical form of the elements in the unravelled sequence $c_j$ (in Algorithm~\ref{algo:GeneratingIndex} we gave an algorithm to generate finitely many elements). However, it is still possible to derive some results on magnitude of $c_j$. To this end, we need the following concept of divisor function.

\begin{definition}
\label{def:divisorfunction}
We use $\tau_D(\cdot):\mathbb{N}^+ \rightarrow \mathbb{N}^+$ to denote the $D$-th divisor function, which counts the number of divisors of an integer (including 1 and the number itself). Formally,
\begin{equation}
    \tau_D(n) =\sum_{\substack{(\mathbf{j}^1,...,\mathbf{j}^D)\in (\mathbb{N}^+)^D\\ \prod_{k=1}^D \mathbf{j}^k =n}} 1
\end{equation}
\end{definition}

We note that $\tau_D$ distinguishes the order of factorization: For example $\tau_2(4) = 3$ because there are $3$ ways to write $4$ as a product of $2$ numbers: $4 = 1\times 4 = 4\times 1 = 2 \times 2$. 

The first several elements in $c_{j}$ are $1,2,2,3,3,4,4,4,....$. As our readers may notice, each natural number $n$ shows up exactly $\tau_2(n)$ times: if we know (averagely) how many ways there are to factorize an integer, we can sketch the general magnitude of the unravelled sequence as well. The following lemma formalizes such an idea:

\begin{lemma}
\label{lemma:numbertheory}
Define $c_{\mathbf{j}} = \prod_{k = 1}^D \mathbf{j}^k$ as a function on the $D$-tuple $\mathbf{j} = (\mathbf{j}^1,...,\mathbf{j}^D)\in (\mathbb{N}^+)^D$. Let $c_j$ be the $c_{\mathbf{j}}$-unravelling sequence of $c_{\mathbf{j}}$ (see definition in Section~\ref{section:unravelling}). Then we know its asymptotic magnitude:
\begin{equation}
    c_j = \Theta\left(j \log^{-(D-1)}j\right)
\end{equation}
\end{lemma}

\begin{proof}
All the elements of $c_j$ are positive integers since they are products of positive integers. And every positive integer shows up in $c_j$ at least once. We also observe that there are repeated elements in $c_j$: For any positive integer $m$, it shows up exactly $\tau_D(m)$ times in the sequence $c_j$.

To determine the increase rate of $c_j$, it is enough to determine the largest $b_j$ such that
\begin{equation}
    \sum_{m=1}^{b_j}\tau_D(m)\leq j.
\end{equation}
The sequence of interest, $c_j$, increases at the same rate as $b_j$.
To quantify the summation on the LHS, we need to use the following result from number theory:
\begin{equation}
\label{eq:tauD}
    \sum_{m=1}^{x} \tau_D(m) =  \frac{\log^{D-1} x}{(D-1) !} x + O\left(x\log^{D-2} x\right),
\end{equation}
where the big $O$ notation is for $x\rightarrow\infty$. If we divide both sides by $x$, then we can read out \eqref{eq:tauD} as: averagely, there are $log^{D-1} x$ ways to factorize a natural number into a product of $D$ natural numbers. This result has been established in the literature of number theory, we give more discussion and references in Appendix~\ref{section:NumberTheoryResults}. For the special case when $D = 2$, there are available sharper results, e.g. Theorem~3.2 \cite{tenenbaum2015introduction}. 

Let $b_j = \lfloor (D-1)!j\log^{-(D-1)}j\rfloor$, then apply \eqref{eq:tauD}
\begin{equation}
\begin{aligned}
     \sum_{m=1}^{b_j} \tau_D(m) &= b_j\log^{D-1} b_j + O\left(b_j\log^{D-2} b_j\right) \\
    & = \Theta\left(  j(\log j)^{-(D-1)} \log^{D-1} \left(j(\log j)^{-(D-1)}\right)\right) = \Theta(j)
\end{aligned}
\end{equation}
And it is direct to check if $b_j= q_j j \log^{-(D-1)} j$ for any positive $q_j\rightarrow\infty$, $b_j\log^{D-1} b_j$ would diverge faster than $j$. So we know the largest $b_j$ we can take is of the order $j\log^{-(D-1)} j$, which concludes our proof. 
\end{proof}

\begin{corollary}
\label{corollary:unravelling}
Let $c_{\mathbf{j}} = \prod_{k = 1}^D (\mathbf{j}^k)^s$ be a function defined on the $D$-tuple $\mathbf{j} = (\mathbf{j}^1,...,\mathbf{j}^D)\in (\mathbb{N}^+)^D$ for some $s > 0$. Let $c_j$ be the $c_{\mathbf{j}}$-unravelling sequence of $c_{\mathbf{j}}$. Then we know
\begin{equation}
    c_j = \Theta\left(\left(j \log^{-(D-1)}j\right)^s\right)
\end{equation}
\end{corollary}

\begin{proof}
The first several elements in $c_j$ are: $1^s, 2^s, 2^s, 3^s,3^s,...$. We can apply the same proof of Lemma~\ref{lemma:numbertheory}.
\end{proof}

The next theorem is the main result in this section, which uses Lemma~\ref{lemma:numbertheory} and Corollary~\ref{corollary:unravelling}.

\begin{theorem}
\label{theorem:threeellipsoids}
Let $W(s,Q,\{\psi_{\mathbf{j}}\})$ be the multivariate product Sobolev ``ellipsoid":
\begin{equation}
\label{eq:multiellipsoid}
    W\left(s,Q,\{\psi_{\mathbf{j}}\}\right) = \left\{f = \sum_{\mathbf{j}\in (\mathbb{N}^+)^D} \beta_{\mathbf{j}}\psi_{\mathbf{j}}, \text{ for some }\beta_{\mathbf{j}}\in \mathbb{R}\ \mid\ \sum_{\mathbf{j}\in (\mathbb{N}^+)^D} c_{\mathbf{j}}^{2s}\beta_{\mathbf{j}}^2 \leq Q^2\right\},
\end{equation}
where $c_{\mathbf{j}} = \prod_{k = 1}^D \mathbf{j}^k$ for $\mathbf{j} = (\mathbf{j}^1,...,\mathbf{j}^D)\in (\mathbb{N}^+)^D$. Denote $(\psi_j)$ to be the $c_{\mathbf{j}}$-unravelling of $\{\psi_{\mathbf{j}}\}$.

Then there exists two constants $C_i(s,D)$, $i\in\{1,2\}$ such that
\begin{equation}
\label{eq:threeellipsoids}
\begin{aligned}
               &\left\{f = \sum_{j=1}^{\infty} \beta_j\psi_j \text{, for some }\beta_{j}\in \mathbb{R}\ \mid\ \sum_{j=1}^{\infty} \left(\frac{j}{\log^{D-1}j \vee 1} \right)^{2s}\beta_j^2 \leq C_1(s,D) Q^2\right\}\\
        \subset &  \quad W(s,Q,\{\psi_{\mathbf{j}}\})\\
        \subset &\left\{f = \sum_{j=1}^{\infty} \beta_j\psi_j\text{, for some }\beta_{j}\in \mathbb{R}\ \mid\ \sum_{j=1}^{\infty} \left(\frac{j}{\log^{D-1}j\vee 1} \right)^{2s}\beta_j^2 \leq C_2(s,D) Q^2\right\}.
\end{aligned}
\end{equation}
\end{theorem}

In language, Theorem~\ref{theorem:threeellipsoids} states that: The multivariate function space $W(s,Q,\{\psi_{\mathbf{j}}\})$ can be sandwiched between two formally simpler function spaces. These ``bread" function spaces in \eqref{eq:threeellipsoids} are still multivariate function spaces, but the basis functions $(\psi_j)$ are listed in a sequence. In contrast, $W(s,Q,\{\psi_{\mathbf{j}}\})$ has basis functions indexed by $D$-tuples.

\begin{proof}
The multivariate ellipsoid $W(s,Q,\{\psi_{\mathbf{j}}\})$ is exactly the same space as:
\begin{equation}
\label{eq:anotherway}
    \left\{f = \sum_{j=1}^{\infty} \beta_j \psi_j \in L^2(\nu)\ \mid\ \sum_{j=1}^{\infty} c_j^{2s}\beta_j^2 \leq Q^2\right\},
\end{equation}
where $c_j, \beta_j, \psi_j$ are the $c_{\mathbf{j}}$-unravelling of $c_{\mathbf{j}}, \beta_{\mathbf{j}}, \psi_{\mathbf{j}}$, respectively. According to Corollary~\ref{corollary:unravelling}, $c_j$ is asymptotically of the same order as $\left(j \log^{-(D-1)}j\right)^{2s}$ as $j\rightarrow\infty$. Define $b_j = \left(\frac{j}{\log^{D-1}j \vee 1} \right)^{2s}$, then we know that there exist constants $C_1, C_2$ (that only depends on $s,D$) such that $C_1 b_j \leq c_j \leq C_2 b_j$ for all $j \in \mathbb{N}^+$. Plugging this in \eqref{eq:anotherway} will conclude our proof.
\end{proof}

\subsection{Approximation in Dense Tensor Product Models}
In this section, we will use the results in  Theorem~\ref{theorem:threeellipsoids} to derive some approximation results that are crucial to understand the performance of sieve estimators. Let's denote the three function spaces in \eqref{eq:threeellipsoids} as $W_1, W_2$ and $W_3$ ($W_1 \subset W_2 \subset W_3$). To study the problem of approximation/estimation functions in $W_2$, it is equivalent -- up to a constant -- to study the corresponding problems in $W_1$ or $W_3$. The regression problem under the assumption $f^0\in W_2$ is easier than assuming $f^0\in W_3$ but harder than $f^0 \in W_1$. Therefore the generalization error of any estimators for truth $f^0\in W_2$ should be of the same order as $f^0 \in W_1\text{ or }W_3$. Similar statements also hold for minimax rates analysis. Ellipsoids related to a (univariate) series $c_j$ can be treated much more directly than ones related to $D$-tuple $c_{\mathbf{j}}$. For readers who are familiar with classical projection estimators (e.g. \cite{introtononpara}), the following approximation results may appear very familiar.

\begin{lemma}
\label{lemma:approximationinellipsoid}
Suppose function $f^*$ has the expansion $ f^* = \sum_{j=1}^{\infty} \beta^*_j\psi_j$ with respect to a set of $\nu$-orthonormal system, i.e. $\langle \psi_j, \psi_i\rangle_{L^2(\nu)} = \delta_{ij}$. Assume $\|\psi_j\|\leq M$ for all $j$. If the expansion coefficients satisfy the following ellipsoid-type condition:
\begin{equation}
    f^* = \sum_{j=1}^{\infty} \beta^*_j\psi_j \in \left\{f = \sum_{j=1}^{\infty} \beta_j\psi_j \in L^2(\nu)\ \mid\ \sum_{j=1}^{\infty} \left(\frac{j}{\log^{D-1}j \vee 1} \right)^{2s}\beta_j^2 \leq Q^2\right\},
\end{equation}
with some $s>1/2$. Then the sequence of functions 
\begin{equation}
    f^*_{n} = \sum_{j=1}^{J_n}\beta^*_{nj}\psi_j\ \text{with}\ J_n = \lfloor (\log^{D-1}n)^{\frac{2s}{2s+1}} n^{\frac{1}{2s+1}} \rfloor, \quad n=2,3,...
\end{equation}
satisfy:
\begin{itemize}
    \item There is a constant $c(M,s,D,f^*)$, for any $n$:
    \begin{equation}
        \|f^*_n\|_{\infty} \leq c(M,s,D)
    \end{equation}
    \item For any measure $\rho_X$ that is absolute continuous to $\nu$ with a bounded density:
    \begin{equation}
    \begin{aligned}
              \|f_n^* - f^*\|^2_{2,\rho_X} = \int_{\mathcal{X}}\left( f_n^*(\tau) - f^*(\tau)\right)^2d\rho_X\\
        \leq  c(M, s,D,\rho_X) \left(\frac{\log^{D-1}n}{n}\right)^{\frac{2s}{2s+1}}
    \end{aligned}
\end{equation}
\end{itemize}
\end{lemma}
\begin{proof}
\begin{itemize}
    \item We first proof the uniform bound in the $\|\cdot\|_{\infty}$-norm. According to our discussion Appendix~\ref{app:MultivariateRKHS}, a Sobolev-ellipsoid can be seen as a ball in an RKHS. That is, the functions $f^*,f^*_n$ all belong to an RKHS with reproducing kernel
    \begin{equation}
        k(s,t) = \sum_{j= 1}^{\infty} \lambda_j \psi_j(s)\psi_j(t),
    \end{equation}
    where $\lambda_j = \left(\frac{\log^{D-1}j\vee 1}{j}\right)^{2s}$. Denote the RKHS inner product as $\langle \cdot, \cdot \rangle_k$:
    \begin{equation}
        \begin{aligned}
            \|f_n^*\|_{\infty} 
            & = \sup_{x\in\mathcal{X}} f_n^*(x) = \sup_{x\in \mathcal{X}}\langle f_n^*, k(x,\cdot)\rangle_{k}\\
            &\leq \|f_n^*\|_{k} \sup_{x\in\mathcal{X}}\|k(x,\cdot)\|_{k}\\
            &\stackrel{(1)}{\leq} \|f^*\|_{k} c(M,s,D).
        \end{aligned}
    \end{equation}
    In step (1), we need the explicit representation of the RKHS norm (Theorem~\ref{theorem:equivalentdef}). The RKHS norm of kernel $k$ (centered at $x$) is 
    $$\|k(x,\cdot)\|_{k} = \sum_{j=1}^{\infty} (\lambda_j \psi_j(x))^2/\lambda_j \leq M^2 \sum_{j=1}^{\infty}\lambda_j = c(M,s,D)$$
    \item Next we proof the bound in $\rho_X$-2-norm:
    \begin{equation}
        \begin{aligned}
        \|f_n^* - f^*\|^2_{2,\rho_X} 
        &\leq U \|f_n^* - f^*\|^2_{2,\nu} = \sum_{J_n + 1}^{\infty} (\beta_j^*)^2\\
        & \leq \lambda_{J_n}\sum_{J_n + 1}^{\infty} (\beta_j^*)^2/\lambda_j \leq \lambda_{J_n} Q^2
        \end{aligned}
    \end{equation}
    We just need to determine the magnitude of $\lambda_{J_n}$:
    \begin{equation}
        \begin{aligned}
            \lambda_{J_n} & \leq  c(J_n)^{-2s}(\log^{D-1}J_n)^{2s}\\
            & = c\left[(\log^{D-1}n)^{\frac{2s}{2s+1}} n^{\frac{1}{2s+1}}\right]^{-2s}
            \left\{\log^{D-1}\left[(\log^{D-1}n)^{\frac{2s}{2s+1}} n^{\frac{1}{2s+1}}\right]\right\}^{2s}\\
            & \leq c(s,D) n^{-\frac{2s}{2s+1}}(\log^{D-1}n)^{-\frac{4s^2}{2s+1} + 2s} = c(s,D)\left(\frac{\log^{D-1}n}{n}\right)^{\frac{2s}{2s+1}},
        \end{aligned}
    \end{equation}
    which concludes our proof.
\end{itemize}
\end{proof}

\subsection{Approximation in Sparse Tensor Product Models}
\label{app:appinsparsemulti}

In the last section we investigated the approximation error under the dense tensor product models. In this section we will switch to the sparse, high dimensional setting. 

Now we present some more general conditions on the product basis and sparse nonparametric models. This can be seen as a generalization of (\textbf{SpTB}) and (\textbf{SpS}) conditions in the main text. 

\begin{itemize}
    \item[(\bf{SpTB'})] Let $\phi_j$ be an orthonormal system of univariate functions. The orthonomality is defined with respect to a measure $\nu$ defined on $\mathcal{X}$, that is, $\langle \phi_i, \phi_j\rangle_{L^2(\nu)} = \delta_{ij}$. Here the domain $\mathcal{X}$ is a compact subset of $\mathbb{R}$. Assume $\phi_1 = 1$, $\|\phi_j\|_{\infty} \leq M$ for all $j = 1,2,...$. Consider their natural $d$-dimensional product extension $\psi_{\mathbf{j}}(\mathbf{x}) = \prod_{k=1}^{d}\phi_{\mathbf{j}^k}(\mathbf{x}^k)$, denote $\psi_j$ to be their $c_{\mathbf{j}}$-unravelling sequence. The unravelling rule $c_{\mathbf{j}}$ is defined as
    \begin{equation}
        c_{\mathbf{j}} = \left\{ \begin{array}{cc}
            \prod_{k=1}^d \mathbf{j}^k &  \text{ , if at most }D' \text{ entries of } \mathbf{j} \text{ are greater than }1 \\
            \infty & \text{ otherwise}
        \end{array}\right.
    \end{equation}
\end{itemize}

\begin{itemize}
    \item[(\bf{SpS'})] There exists a $D$-variate function $f^*:\mathcal{X}^D\rightarrow \mathbb{R}$ such that:
    \begin{enumerate}
       \item There is set of indices $\{k_1,..,k_D\}\subset\{1,2,...,d\}$ such that for any $\mathbf{u}~\in~ \mathcal{X}^d$,
        \begin{equation}
            f^0(\mathbf{u}) = f^*(\mathbf{u}^{k_1}, \mathbf{u}^{k_2}, ...,\mathbf{u}^{k_D}).
        \end{equation}
        \item The function $f^*$ satisfies the following ellipsoid assumption:
\begin{equation}
\label{eq:squareellipsoid}
    f^* \in \left\{f = \sum_{j=1}^{\infty} \beta_j\square_j\ \mid\ \sum_{j=1}^{\infty} \left(\frac{j}{\log^{D-1}j \vee 1} \right)^{2s}\beta_j^2 \leq Q^2\right\}.
\end{equation}
The function sequence $(\square_j)$ is the $\triangle_{\mathbf{j}}$-unravelling of $\square_{\mathbf{j}} = \prod_{k=1}^D \phi_{\mathbf{j}_k}(\mathbf{u}^k)$, $\mathbf{j}\in(\mathbb{N}^+)^D$. And the unravelling rule is defined by $\triangle_{\mathbf{j}} = \prod_{k=1}^{D}\mathbf{j}^k$.
    \end{enumerate}
\end{itemize}

The assumption 1 of (SpS') is a feature sparsity condition. Although $f^0$ formally is a function of $d$-dimensional vector $\mathbf{x}$ ($d$ can be large), this assumption states that it can be well-described using a small subset of the dimensions of $\mathbf{x}$ (specifically, we assume it depends on $D$ out of the $d$ dimensions).

The requirement 2 of (SpS') is in nature a smoothness assumption, but expressed in a basis expansion/Sobolev ellipsoid fashion. The basis functions $\square_{\mathbf{j}}$ and unravelling rules $\triangle_{\mathbf{j}}$ only engage with the ``correct features". According to Lemma~\ref{lemma:approximationinellipsoid}, if we use the first $J_n^{\text{oracle}} = \lfloor (\log^{D-1}n)^{\frac{2s}{2s+1}} n^{\frac{1}{2s+1}} \rfloor$ functions of $\square_j$, we can construct a sequence of approximation functions $f^{\text{oracle}}_n = \sum_{j=1}^{J_n}\beta^{\text{oracle}}_{nj} \square_j$ of $f^*$ that satisfy 

$$\|f_n^{\text{oracle}} - f^*\|^2_{2,\rho_X} \leq  c(s,D,\rho_X, f^*) \left(\frac{\log^{D-1}n}{n}\right)^{\frac{2s}{2s+1}}.$$

However, in real-world problems, we are unfortunately do not have \textit{a priori} accessible information of which $D$ dimensions of $\mathbf{x}$ are effectively associated with the outcome $Y$. We end up cannot use the ``optimal" basis $\square_j$ that only depends on the $D$ relevant dimensions. The basis functions we use in \eqref{eq:LASSOproblem} take the form of $\psi_{\mathbf{j}} = \prod_{k=1}^d \phi_{\mathbf{j}^k}(\mathbf{x}^k)$, involving $d$ univariate functions as described in (SpTB'). We are interested in how many functions we need to include in the sequence of $\psi_j$, such that we can achieve the same approximation error as $f_n^{\text{oracle}}$. The following Lemma tells us this number is exponential in the intrinsic dimension $D$ (which we treat as a ``fixed" number) but only polynomial in the ambient dimension $d$ (which may formally ``increase" with the sample size $n$).

\begin{lemma}
\label{lemma:approximationsparsemodel}
Assume $f^0$ satisfies the condition (SpS'). Denote $\{\psi_j\}$ to be the sequence of product basis functions in (SpTB'). If the working dimension $D'$ in (SpTB') is greater or equal to the intrinsic dimension $D$ in (SpS'), then:

\begin{itemize}
    \item The true regression function $f^0$ can be expanded with respect to $\psi_j$ as well, that is,
    \begin{equation}
    \label{eq:basisis rich}
        f^0 = \sum_{j=1}^{\infty}\beta_{j}^0\psi_j, \text{ for }\beta^0_j\in \mathbb{R}.
    \end{equation}
    \item There exists a sequence of functions $f_{\beta_n^0}= \sum_{j = 1}^{J_n} \beta^0_{nj}\psi_j$ with $J_n \leq  C(s,D)d^{D'}n^{1/(2s+1)} \log^{D'-1} n$ such that
    \begin{equation}
         \|f_{\beta_n^0}\|_{\infty} \leq c(M,s,D, f^0)
    \end{equation}
    and
    \begin{equation}
      \|f_{\beta_n^0} - f^0\|^2_{2,\rho_X} \leq  c(s,D,\rho_X, f^0) \left(\frac{\log^{D-1}n}{n}\right)^{\frac{2s}{2s+1}}.
\end{equation}

\end{itemize}
\end{lemma}
\begin{proof}
We introduce the mapping $1_{d\rightarrow D}:\mathbb{R}^d\rightarrow \mathbb{R}^D$ that only keeps the relevant dimensions of a feature $\mathbf{x}$:
\begin{equation}
    1_{d\rightarrow D}(\mathbf{x}) = (\mathbf{x}^{k_1},...,\mathbf{x}^{k_D}),
\end{equation}
where $k_1,...,k_D$ are the dimension indices defined in (SpS'). By our assumptions in (SpS'), the true regression function can be written as:
\begin{equation}
    f^0(\mathbf{x}) = f^*(1_{d\rightarrow D}(\mathbf{x})) = \sum_{j=1}^{\infty} \beta^*_j\square_j(1_{d\rightarrow D}(\mathbf{x})) = \sum_{j=1}^{\infty} \beta^*_j\square_j\circ 1_{d\rightarrow D}(\mathbf{x})
\end{equation}
Each of the basis functions above, $\square_j\circ 1_{d\rightarrow D}$, varies at most in $D$ dimensions. The function set $\{\psi_j\}$ in (TB') includes all the function product functions varying in at most $D'$ dimensions. Since $\square_j\circ 1_{d\rightarrow D}$ are also product functions, we conclude $\{\square_j\circ 1_{d\rightarrow D}, j\in\mathbb{N}\}\subset \{\psi_j, j\in\mathbb{N}\}$. Therefore $f^0$ also has the expansion with respect to $\psi_j$ as in \eqref{eq:basisis rich}.


Approximating $f^0$ (or equivalently, $f^*$), using the ``correct" basis $\square_j$ in the ellipsoid assumption \eqref{eq:squareellipsoid}, is already studied in Lemma~\ref{lemma:approximationinellipsoid}. We know that we need the first $S_n = (\log^{D-1}n)^{\frac{2s}{2s+1}} n^{\frac{1}{2s+1}}$ basis in $\{\square_j\}$ to achieve the desired approximation error. We know the first $S_n$ elements can be included in the following set
\begin{equation}
\label{eq:firsthowevermanysquare}
    \left\{\square_{\mathbf{j}}, \mathbf{j}\in(\mathbb{N}^+)^D\ \mid \ c_{\mathbf{j}} = \prod_{k=1}^D \mathbf{j}^k \leq C(s,D) n^{\frac{1}{2s+1}} \log^{-\frac{D-1}{2s+1}} n\right\}.
\end{equation}
To see this, we need to apply the number theory results we used to establish the equivalence between ellipsoids. Use the $T_D$ notation in Lemma~\ref{def:TD}, according to Lemma~\ref{lemma:numbertheory} we know that
\begin{equation}
    T_D\left(C(s,D) n^{\frac{1}{2s+1}} \log^{-\frac{D-1}{2s+1}} n\right) = C(s,D)n^{\frac{1}{2s+1}} \log^{\frac{(D-1)2s}{2s+1}}n + \text{ lower order terms}.
\end{equation}

Therefore, to approximate $f^0$ well, we need to choose $J_n$ large enough so that all the functions below are included:
\begin{equation}
    \left\{\psi_{\mathbf{j}}, \mathbf{j}\in(\mathbb{N}^+)^d\ \mid \ c_{\mathbf{j}} = \prod_{k=1}^d \mathbf{j}^k \leq C(s,D) n^{\frac{1}{2s+1}} \log^{-\frac{D-1}{2s+1}} n \text{ and at most } D \text{ of }\mathbf{j}^k \text{ are greater than }1 \right\}.
\end{equation}
Compared with \eqref{eq:firsthowevermanysquare}, we changed the index dimension from $D$ to $d$. But these two sets corresponds to the same collection of functions. By our assumption that $D'>D$, we only need to select $J_n$ large enough so that the following basis functions are all included:
\begin{equation}
\label{eq:howevermanypsi}
\begin{aligned}
&\left\{\psi_{\mathbf{j}}, \mathbf{j}\in(\mathbb{N}^+)^d\ \mid \ c_{\mathbf{j}} = \prod_{k=1}^d \mathbf{j}^k \leq C(s,D) n^{\frac{1}{2s+1}} \log^{-\frac{D-1}{2s+1}} n\text{ and at most }D'\text{ of }\mathbf{j}^k \text{ are greater than 1}\right\}\\
& = \bigcup_{m = 1}^{\left\lfloor C(s,D) n^{\frac{1}{2s+1}} \log^{-\frac{D-1}{2s+1}} n \right\rfloor } \left\{\psi_{\mathbf{j}}\ \mid \ c_{\mathbf{j}} = \prod_{k=1}^d \mathbf{j}^k = m\text{ and at most }D'\text{ of }\mathbf{j}^k \text{ are greater than 1}\right\}
\end{aligned}
\end{equation}

How many elements are there in \eqref{eq:howevermanypsi}? We give the following bound:
\begin{equation}
    \begin{aligned}
        & \# \text{ of elements in \eqref{eq:howevermanypsi}}  \leq\\
       & \underbrace{\sum_{m=1}^{C(s,D) n^{\frac{1}{2s+1}} \log^{-\frac{D-1}{2s+1}} n}}_{\text{consider all the }{\mathbf{j}} \text{ whose product is }m}  \underbrace{C^{d}_{D'}}_{\text{choose }D'\text{ dimensions}} \cdot\qquad \underbrace{\tau_{D'}(m)}_{\text{factorize }m \text{ into a product of }D' \text{numbers}}\\
& \leq C^{d}_{D'} T_{D'}(C(s,D) n^{\frac{1}{2s+1}} \log^{-\frac{D-1}{2s+1}} n) \stackrel{(1)}{\leq} C(s,D,D') d^{D'} n^{\frac{1}{2s+1}} \log^{D'-1} n.
    \end{aligned}
\end{equation}
In (1) we used Lemma~\ref{lemma:numbertheory} and the well-known bound on the binomial coefficients $C^{d}_{D'} \leq C(D') d^{D'}$. 
Unravel the functions indexed by $\mathbf{j}$ will give us at most the first $C(s,D, D') d^{D'} n^{\frac{1}{2s+1}} \log^{D'-1} n$ elements in $\psi_j$. To achieve the desired approximation ability, we do not need to go further than the first $J_n = C(s,D, D') d^{D'} n^{\frac{1}{2s+1}} \log^{D'-1} n$ elements in $\psi_j$.
\end{proof}

\newpage

\section{Theoretical Guarantee of Penalized Sieve Estimators}

To present the statistical guarantee of $l_1$-penalized sieve estimators, we are going to derive the following results in sequence:
\begin{itemize}
    \item Some nonparametric oracle inequalities to control the training error of the estimators and the deviation of the estimated regression coefficients (Corollary~\ref{corollary:cleanversion1}).
    \item Use the information of the regression coefficients to derive a metric entropy bound on the function space the estimator lies in (Lemma~\ref{lemma:entropyboundonG}).
    \item Control the difference between the training and testing errors of the estimate using results from empirical process theory (Theorem~\ref{th:empiricalprocess}).
\end{itemize}

\subsection{Nonparametric Oracle Inequalities}
We first define the concept of compatibility constant, which is an important component in the oracle inequalities.

\begin{definition}
\label{def:compatibility}
For a given matrix $\Sigma$ of size $J\times J$, constant $L$, and an index set $S\subset \{1,2,...,J\}$, we define the $(\Sigma,L,S)$-compatibility constant $\phi_{\Sigma}(L,S)$ to be
\begin{equation}
\phi^{2}_{\Sigma}(L, S) =\min_{\beta} \left\{\frac{|S|\beta^{\top}\Sigma\beta}{\|\beta_S\|_1^2}:\left\|\beta_{-S}\right\|_{1}\leq L\|\beta_S\|_1\right\},
\end{equation}
where $-S$ is the complementary set of $S$ in $\{1,2,..,J\}$. The notation $\beta_{S}\in\mathbb{R}^J$ is a shorthand for the ``restriction" of a vector $\beta\in\mathbb{R}^J$ on the index set $S$: $(\beta_S)_j = \beta_j$ if $j\in S$, otherwise $(\beta_S)_j = 0$.

\end{definition}

The following oracle inequality is a generalization of Theorem~2.2 in \cite{van2016estimation}. In our case, the true regression function does not have to be linear.

\begin{theorem}
\label{theorem:oraclebasic}
Let $(\psi_j)$ be the unravelled sequence described in (SpTB'). Let $\lambda_{\epsilon}$ be a number satisfying:
\begin{equation}
\sup_{1\leq j\leq J_n}\left|\frac{1}{n}\sum_{i=1}^n\psi_j(\mathbf{X}_i)\epsilon_i\right|\leq \lambda_{\epsilon} 
\end{equation}
Let $0 \leq \delta<1$ and define for $\lambda>\lambda_{\epsilon}>0$:
\begin{equation}
\underline{\lambda} =\lambda-\lambda_{\epsilon}, \bar{\lambda} =\lambda+\lambda_{\epsilon}+\delta \underline{\lambda}, \text{ and } L=\frac{\bar{\lambda}}{(1-\delta) \lambda}.
\end{equation}
We use $\hat\beta_n = (\beta_1^{PLS}, ...,\beta_{J_n}^{PLS})^{\top}$ to denote the minimizer of the penalized problem \eqref{eq:LASSOproblem}. For a $\beta\in\mathbb{R}^{J_n}$, we define a related function $f_{\beta}$ as $f_{\beta} = \sum_{j=1}^{J_n}\beta_j\psi_j$.

Then for any $\beta \in \mathbb{R}^{J_n}$ and any set $S\subset \{1,2,...,J_n\}$:
\begin{equation}
2 \delta \underline{\lambda}\|\hat{\beta}-\beta\|_{1}+\left\|f_{\hat\beta_n} - f^0\right\|_{n}^{2} 
\leq\left\|f_{\beta}-f^0\right\|_{n}^{2}+\frac{\bar{\lambda}^{2}|S|}{\phi^2_{\hat\Sigma}(L, S)}+4 \lambda\left\|\beta_{-S}\right\|_{1}
\end{equation}
where $\phi^2_{\hat\Sigma}(L, S)$ is the $(\hat\Sigma, L,S)$-compatibility constant and the $\hat\Sigma$ is the empirical covariance matrix: $\hat \Sigma_{ij} = \frac{1}{n}\sum_{k=1}^{n}\psi_i(\mathbf{X}_k)\psi_j(\mathbf{X}_k)$.

\begin{proof}
We denote $2\square = \|f_{\hat\beta_n} - f^0\|_n^2 - \|f_{\beta} - f^0\|_n^2 + \|f_{\hat\beta_n} - f_{\beta}\|_n^2$. The empirical norm $\|\cdot\|_n$ can also be written in a matrix form, for example:
\begin{equation}
    \begin{aligned}
        \|f_{\hat\beta_n} - f^0\|_n^2 = & \frac{1}{n}\sum_{i=1}^n \left( f_{\hat\beta_n}(\mathbf{X}_i) - f^0(\mathbf{X}_i)\right)^2\\
        & = \frac{1}{n}\sum_{i=1}^n \left( \sum_{j=1}^{J_n} \beta_j^{PLS}\psi_j(\mathbf{X}_i) - f^0(\mathbf{X}_i)\right)^2\\
        & = \langle \hat\Psi \hat\beta_n - f^0(X), \hat\Psi \hat\beta_n - f^0(X)\rangle/n.
    \end{aligned}
\end{equation}

The design matrix, $\hat \Psi$, has entries $\hat\Psi_{i,j} = \psi_j(\mathbf{X}_i)$. And $f^0(X)\in \mathbb{R}^n$ is the evaluation vector of $f^0$ at $n$ features vectors $\{\mathbf{X}_i\}_{i=1}^n$. Similar to the proof in the literature, we consider two cases of $\square$:

\begin{itemize}
    \item If $\square \leq -\delta \underline{\lambda} \|\hat\beta_n - \beta\|_1 + 2\lambda \|\beta_{-S}\|_1$.
    Then we have 
    \begin{equation}
        \begin{aligned}
            2\delta \underline{\lambda} &\|\hat\beta_n - \beta\|_1 + \|f_{\hat\beta_n} - f^0\|_n^2\\
            & = 2\delta \underline{\lambda} \|\hat\beta_n - \beta\|_1 + 2\square + \|f_{\beta} - f^0\|_n^2 - \|f_{\hat\beta_n} - f_{\beta}\|_n^2\\
            & \leq \|f_{\beta} - f^0\|_n^2 - \|f_{\hat\beta_n} - f_{\beta}\|_n^2 + 4\lambda \|\beta_{-S}\|_1\\
            & \leq \|f_{\beta} - f^0\|_n^2 + 4\lambda \|\beta_{-S}\|_1
        \end{aligned}
    \end{equation}
    \item In the case when $\square > -\delta \underline{\lambda} \|\hat\beta_n - \beta\|_1 + 2\lambda \|\beta_{-S}\|_1$, we start with the following \textit{two point inequality} (Lemma~6.1 in \cite{van2016estimation}):
    \begin{equation}
    \label{eq:twopoint}
        \langle \hat\Psi (\beta - \hat\beta_n), Y - \hat\Psi \hat\beta_n \rangle / n \leq \lambda\|\beta\|_1 - \lambda \|\hat\beta_n\|_1
    \end{equation}
    Use the results in the beginning of this proof, we know $\square$ can be expanded as:
    \begin{equation}
        \square = \langle \hat\Psi \hat\beta_n,\hat\Psi \hat\beta_n \rangle/n 
        - \langle \hat\Psi \hat\beta_n,f^0(X) \rangle/n 
        + \langle \hat\Psi \beta,f^0(X) \rangle/n 
        - \langle \hat\Psi \hat\beta_n,\hat\Psi \beta \rangle/n
    \end{equation}
    Then eq \eqref{eq:twopoint} implies that:
    \begin{equation}
        \begin{aligned}
            \square  & \leq \langle \hat\Psi \hat\beta_n, \mathbf{\epsilon} \rangle/n
            - \langle \hat\Psi \beta, \mathbf{\epsilon} \rangle/n + \lambda \|\beta\|_1 - \lambda\|\hat\beta_n\|_1,
        \end{aligned}
    \end{equation}
    The $\epsilon$ vector stores the noise variables: $\epsilon_i = Y_i - f^0(\mathbf{X}_i)$. The rest of the proof follows line by line as that of Theorem~2.2 in \cite{van2016estimation} (page 21), replacing the $(\hat{\beta}-\beta)^{\top} \hat{\Sigma}\left(\hat{\beta} -\beta^{0}\right)$ term there by $\square$. 
\end{itemize}

\end{proof}
\end{theorem}

The following lemmas tell us the random compatibility constant $\phi_{\hat\Sigma}(L, S)$ is bounded away from zero with high probability.

\begin{lemma}
\label{lemma:populationcompatibility}
Let $\Sigma$ be the population $J_n \times J_n$ covariance matrix $\Sigma_{ij} = E[\psi_i(\mathbf{X})\psi_j(\mathbf{X})]$, where $(\psi_j)$ is the unravelled function sequence defined in (SpTB'). Assume the feature density function $p_{X}(\mathbf{x}) = d\rho_X/d\nu^{d} \geq u > 0$ is bounded away from $0$. Here $\nu^d$ is the $d$-dimension product measure of $nu$.

Then we know $\Sigma$ has a compatibility constant $\phi^2_{\Sigma}(L,S) \geq u$ for any $L$ and $S$.
\end{lemma}

\begin{proof}
For any $\beta \in \mathbb{R}^{J_n}$:
\begin{equation}
    \begin{aligned}
        \beta^{\top}\Sigma\beta & = \sum_{1\leq i,j \leq J_n} \beta_i \beta_j E[\psi_i(\mathbf{X})\psi_j(\mathbf{X})]
         = E\left[\left(\sum_{j=1}^{J_n} \psi_j(\mathbf{X})\beta_j\right)^2\right]\\
        &\geq u\int \left(\sum_{j=1}^{J_n} \psi_j(\mathbf{x})\beta_j\right)^2 d\mathbf{x}
        \stackrel{(1)}{=} u\|\beta\|_2^2.
    \end{aligned}
\end{equation}
In step (1) we used the orthonomality of $\psi_j$ stated in (SpTb'). At the same time we have $\|\beta_S\|_1^2\leq \|\beta\|_2^2|S| $. Checking the definition of compatibility (Definition~\ref{def:compatibility}), we conclude for any $L,S$, the matrix $\Sigma$ has an uniform compatibility constant $\phi_{\Sigma}(L,S)$ greater than $\sqrt{u}$ (meaning that this lower bound does not depend on either $L$ or $S$).
\end{proof}

\begin{lemma}
\label{lemma:boundonrandomcompatibilityconstant}
Under the same condition as in Lemma~\ref{lemma:populationcompatibility}, we know the empirical matrix $\hat\Sigma$ has a compatibility constant $\phi^2_{\hat\Sigma}(L,S)\geq u/2$, with probability at least $1-J_n^2\exp(-na^2/2M^{4D'})$, $a = u(L+1)^{-2}|S|^{-1}/2$, .

Recall that $u$ is the lower bound on the density and $M$ is the bound on $\|\phi_j\|_{\infty}$, $d$ is the ambient dimension of the feature $\mathbf{X}_i$.
\end{lemma}
\begin{proof}
We first consider the difference between two quadratic forms related to the two covariance matrices:
\begin{equation}
\label{eq:1511}
    | \beta^{\top}\hat\Sigma\beta - \beta^{\top}\Sigma\beta|  = \left|\sum_{1\leq i,j \leq J_n} \beta_i\beta_j(\hat\Sigma_{ij} - \Sigma_{ij}) \right| \leq \|\beta\|_1^2 \|\hat\Sigma - \Sigma\|_{\infty}
\end{equation}
By the definition of $\phi_{\Sigma}(L,S)$, for any $\beta$ such that $\|\beta_{-S}\|_1 \leq L\|\beta_S\|_1$, we have
\begin{equation}
    \|\beta\|_1 \leq (L+1)\|\beta_S\|_1 \leq (L+1)\sqrt{|S|\beta^{\top}\Sigma\beta}/\phi_{\Sigma}(L,S)
\end{equation}
Plug this into \eqref{eq:1511}:
\begin{equation}
    \begin{aligned}
        \mid \beta^{\top}\hat\Sigma\beta - \beta^{\top}\Sigma\beta\mid  &\leq (L+1)^2\|\hat\Sigma-\Sigma\|_{\infty}|S|\beta^{\top}\Sigma\beta/\phi_{\Sigma}^2(L,S)\\
        \Longleftrightarrow \left| \frac{\beta^{\top}\hat\Sigma\beta}{\beta^{\top}\Sigma\beta} - 1\right| &\leq (L+1)^2\|\hat\Sigma-\Sigma\|_{\infty}|S|/\phi^2_{\Sigma}(L,S)
    \end{aligned}
\end{equation}
By a typical application of Hoeffding's inequality (every entry in $\hat \Sigma$ is a bounded random variable), we know with probability at least $1-J^2_n\exp(-na^2/2M^{4D'})$, $a = u(L+1)^{-2}|S|^{-1}/2$
\begin{equation}
    \|\hat\Sigma-\Sigma\|_{\infty} \leq (2(L+1)^2|S|/u)^{-1}
\end{equation}
With the same probability we have
\begin{equation}
    \left| \frac{\beta^{\top}\hat\Sigma\beta}{\beta^{\top}\Sigma\beta} - 1\right| \leq \frac{1}{2}
\end{equation}
Therefore, for all any $\beta$ such that $\|\beta_{-S}\|_1 \leq L\|\beta_S\|_1$:
\begin{equation}
    \frac{|S|\beta^{\top}\hat\Sigma\beta}{\|\beta_S\|_1^2} \geq \frac{|S|\beta^{\top}\Sigma\beta}{2\|\beta_S\|_1^2},
\end{equation}
with high probability. By the definition of the compatibility constant, we can read out
\begin{equation}
    \phi_{\hat\Sigma}^2(L,S) \geq  \phi_{\Sigma}^2(L,S)/2
\end{equation}
which concludes our proof.
\end{proof}

\begin{corollary}
\label{corollary:cleanversion1}
Let $\lambda_{\epsilon} = \sqrt{\frac{2\log(2J_n)}{c(C_{sub},M)n}}$ and assume $\epsilon_i$ to be the uniform sub-Gaussian noise. Then, under the same condition as in Theorem~\ref{theorem:oraclebasic}, for any $\beta\in\mathbb{R}^{J_n}$ whose support is $S\subset \{1,2,...,J_n\}$,  we have 
\begin{equation}
\label{eq:generaloracle}
\lambda_{\epsilon}\|\hat{\beta}_n - \beta\|_{1}+\left\|f_{\hat\beta_n} - f^0\right\|_{n}^{2} 
\leq \frac{3}{2}\left\|f_{\beta}-f^0\right\|_2^{2}+\frac{49\lambda_{\epsilon}^{2}|S|}{2u}
\end{equation}
with probability larger than $1-1/(2J_n) - J_n^2\exp(-na^2/2M^{4D'})-\exp(-cn\|f_{\beta} - f^0\|_2^2/M_0^2) $, and $a = u(L+1)^{-2}|S|^{-1}/2$. The definition of $f_{\beta}$ is stated in Theorem~\ref{theorem:oraclebasic}.
\end{corollary}
\begin{proof}
First we show that for the chosen $\lambda_{\epsilon}$, it can bound the following supremum with high probability:
\begin{equation}
\sup_{1\leq j\leq J_n}\left|\frac{1}{n}\sum_{i=1}^n\psi_j(\mathbf{X}_i)\epsilon_i\right|\leq \lambda_{\epsilon} 
\end{equation}
Since $(\epsilon_i)$ are sub-Gaussian random variables (with a parameter not depending on $\mathbf{X}$), we know there exists a constant $C_{sub}$
\begin{equation}
    \|\epsilon_i\|_{L^{p}}=\left( E|\epsilon_i|^{p}\right)^{1 / p} \leq C_{sub} \sqrt{p} \quad \text { for all } p \geq 1
\end{equation}
For references, see e.g. Proposition 2.5.2 in \cite{vershynin2018high}. The basis functions $\psi_j$ are also uniformly bounded (by $M^{D'}$), so we have
\begin{equation}
    \|\psi_j(\mathbf{X}_i)\epsilon_i\|_{L^{p}}\leq C_{sub}M^{D'} \sqrt{p} \quad \text { for all } p \geq 1.
\end{equation}
This means $\psi_j(\mathbf{X}_i)\epsilon_i$ is also sub-Gaussian. Apply an union bound and Hoeffding's inequality for sub-Gaussian variables (e.g. Theorem~2.6.3 in \cite{vershynin2018high}):
\begin{equation}
    \begin{aligned}
        P\left(\sup_{1\leq j\leq J_n}\left|\frac{1}{n}\sum_{i=1}^n\psi_j(\mathbf{X}_i)\epsilon_i\right| \geq t\right)
        &\leq \sum_{j=1}^{J_n} P\left(\left|\frac{1}{n}\sum_{i=1}^n\psi_j(\mathbf{X}_i)\epsilon_i\right| \geq t\right)\\
        &\leq 2J_n\exp \left(-c(C_{sub},M)n t^{2}\right)
    \end{aligned}
\end{equation}
Take $t = \lambda_{\epsilon} = \sqrt{\frac{2\log(2J_n)}{c(C_{sub},M)n}}$, we know
\begin{equation}
\label{eq:noiseempiricalprocess}
    P\left(\sup_{1\leq j\leq J_n}\left|\frac{1}{n}\sum_{i=1}^n\psi_j(\mathbf{X}_i)\epsilon_i\right| \leq \lambda_{\epsilon}\right) \geq 1-1/(2J_n)
\end{equation}
This is what we claimed in the beginning of the proof.

Next, we bound the difference between $\|f_{\beta} - f^0\|_n^2$ and $\|f_{\beta} - f^0\|_2^2$ for any fixed $f_{\beta}$ satisfying $\|f_{\beta}\|_{\infty} < 2\|f^0\|_{\infty}$. First we note that $(f_{\beta}(\mathbf{X}_i) - f^0(\mathbf{X}_i))^2$ is a bounded variable, therefore it is sub-Gaussian with parameter $9M^2_0$, where $M_0$ is a bound of $\|f^0\|_{\infty}$. The centered version, $(f_{\beta}(\mathbf{X}_i) - f^0(\mathbf{X}_i))^2 - \|f_{\beta} - f^0\|_2^2$ is also sub-Gaussian with parameter $cM^2_0$ (e.g. Lemma~2.6.8 in \cite{vershynin2018high}). Use Hoeffding's inequality again:
\begin{equation}
\begin{aligned}
    P\left(\left|\frac{1}{n}\sum_{i=1}^n(f_{\beta}(\mathbf{X}_i) - f^0(\mathbf{X}_i))^2 - \|f_{\beta} - f^0\|_2^2\right| \geq t\right) \leq \exp(-cnt^2/M_0^2)\\
    \Rightarrow P\left(\left|\frac{\|f_{\beta} - f^0\|_n^2}{\|f_{\beta} - f^0\|_2^2} - 1\right| \geq \frac{1}{2}\right) \leq \exp(-cn\|f_{\beta} - f^0\|_2^2/M_0^2)
\end{aligned}
\end{equation}
We know, with probability larger than $1-\exp(-cn\|f_{\beta} - f^0\|_2^2/M_0^2)$
\begin{equation}
\label{eq:onefunctionratio}
\frac{\|f_{\beta} - f^0\|_n^2}{\|f_{\beta} - f^0\|_2^2} \leq \frac{3}{2}
\end{equation}
Combine \eqref{eq:noiseempiricalprocess}, \eqref{eq:onefunctionratio}, Lemma~\ref{lemma:boundonrandomcompatibilityconstant} and Theorem~\ref{theorem:oraclebasic}, we conclude our proof.
\end{proof}

\subsection{Theoretical Guarantee under Sparse Tensor Product Models}

In this section, we will combine the oracle inequalities developed in the last section with approximation results to derive some performance guarantee of the $l_1$-penalized sieve estimators.

Recall the notation: $d$ is the ambient dimension of feature $\mathbf{X}_i$, $D$ is the number of explanatory features related to the outcome $Y$ (intrinsic dimension), $s$ is the smoothness parameter in (SpS'), and $J_n$ is the number of basis functions in the lasso problem \eqref{eq:LASSOproblem}. The constant $C_{sub}$ is the sub-Gaussian parameter for noise variables, $u$ is the lower bound of the feature density function and $M_0$ is a bound on the $\|\cdot\|_{\infty}$-norm of $f^0$.

\begin{corollary}
\label{corollary:cleanversion2}
Let $f_{\hat\beta_n}$ be the penalized sieve estimate of $f^0$, and $f_{\beta_n^0}$ be the approximate of $f^0$ as in Lemma~\ref{lemma:approximationsparsemodel}. Choose the penalization hyperparameter as $\lambda_{\epsilon} = \sqrt{\frac{2\log(2J_n)}{c(C_{sub},M)n}}$. Under (SpS'), (SpTB') and the boundedness conditions of the feature density function $p_X$, we can show that: 
\begin{equation}
    \begin{aligned}
\left\|f_{\hat\beta_n} - f^0\right\|_{n}^{2} 
\leq c(C_{sub},M,u,f^0)\log(J_n)\left(\frac{\log^{D-1}(n)}{n}\right)^{\frac{2s}{2s+1}}\\
\|\hat{\beta}_n-\beta_n^0\|_{1} \leq c(C_{sub},M,u,f^0) \sqrt{\frac{\log J_n}{n}} n^{\frac{1}{2s+1}} \log^{\frac{2s(D-1)}{2s+1}}(n)
    \end{aligned}
\end{equation}
with probability at least $$1-1/(2J_n) - J_n^2\exp(-na^2/2M^{4D'})-\exp\left(-c(s,D,\rho_X,f^0)(\log n)^{(D-1)2s/(2s+1)} n^{1/(2s+1)}\right), $$ $a = u(L+1)^{-2}|S|^{-1}/2$.
\end{corollary}

\begin{proof}
To get the bounds above, we only need to combine the oracle inequality in Corollary~\ref{corollary:cleanversion1} with the approximation results in Lemma~\ref{lemma:approximationsparsemodel}.

In Lemma~\ref{lemma:approximationsparsemodel}, we discussed that so long as $J_n$ is large enough, we can find a function $f_{\beta_n^0}$ that approximate $f^0$ well enough. Plug in the results of Lemma~\ref{lemma:approximationsparsemodel} into the oracle inequality \eqref{eq:generaloracle}, we have:
\begin{equation}
\label{eq:specialoracle}
\lambda_{\epsilon}\|\hat{\beta}_n - \beta\|_{1}+\left\|f_{\hat\beta_n} - f^0\right\|_{n}^{2} 
\leq  c(s,D,\rho_X, f^0) \left(\frac{\log^{D-1}n}{n}\right)^{\frac{2s}{2s+1}} +\frac{49\lambda_{\epsilon}^{2}|S_n|}{2u}
\end{equation}
Although formally $f_{\beta_n^0}$ is a linear combination of $J_n = C(s,D)d^{D'}n^{1/(2s+1)} \log^{D'-1} n$ basis functions, the size of its support, $|S_n|$, is much smaller than $J_n$. In fact $f_{\beta_n^0}$ only need to engage with the correct features. In Lemma~\ref{lemma:approximationinellipsoid}, we showed that $|S_n|$ can be bounded by $(\log^{D-1}n)^{\frac{2s}{2s+1}} n^{\frac{1}{2s+1}}$. Plug this in the above inequality:

\begin{equation}
\lambda_{\epsilon}\|\hat{\beta}_n - \beta\|_{1}+\left\|f_{\hat\beta_n} - f^0\right\|_{n}^{2} 
\leq  c(s,D,\rho_X, f^0) \left(\frac{\log^{D-1}n}{n}\right)^{\frac{2s}{2s+1}} +c(C_{sub},M, u) \log(J_n) \left(\frac{\log^{D-1}n}{n}\right)^{\frac{2s}{2s+1}}.
\end{equation}
This gives us the results regarding the training error and $l_1$-distance stated in Corollary~\ref{corollary:cleanversion2} (the second term will dominate for large $n$).
\end{proof}

At this point, we already established bounds on the ``training error" bound (expressed in the $\|\cdot\|_n$-norm). However, for most prediction problems we are interested in the ``testing error" (quantified in the $\|\cdot\|_{2,\rho_X}$-norm). For arbitrary flexible estimator, a low training error does not imply a low testing error. However, according to Corollary~\ref{corollary:cleanversion2}, the coefficient $\hat\beta_n$ lives in a small $\|\cdot\|_1$-ball centered around the oracle $\beta_n^0$ with high probability. From this we can also develop some bounds on metric entropy of the space in which $f_{\hat\beta_n}$ takes value. These will in turn link the testing error and the training error together.

Below we will use the concept of metric entropy of a function space. For more comprehensive discussion, see Chapter~2 of \cite{geer2000empirical}.

\begin{definition}
Let $Q$ be a measure on $\mathcal{X}$ and let $\mathcal{G}$ be a function space $\mathcal{G}\subset L_2(\mathcal{X};Q)$. Consider for each $\delta > 0$, a collection of functions $g_1,...,g_N$, such that for each $g\in\mathcal{G}$, there is a $j = j(g) \in\{1,...,N\}$, such that
\begin{equation}
    \left(\int_{\mathcal{X}} (g(x) - g_j(x))^2 d Q(x)\right)^{1/2} \leq \delta.
\end{equation}
Let $N(\delta, \mathcal{G}, Q)$ be the smallest value of $N$ for which such a covering by balls with radius $\delta$ and centers $g_1,...,g_N$ exists. Then $N(\delta, \mathcal{G}, Q)$ is called the covering number (under measure $Q$) and $H(\delta, \mathcal{G}, Q) = \log N(\delta, \mathcal{G}, Q)$ is called the metric entropy of $\mathcal{G}$ (under measure $Q$). 
\end{definition}



One of the function spaces $\mathcal{G}_n$ we are going to consider is
\begin{equation}
\label{eq:defgn}
    \mathcal{G}_n =\mathcal{G}_n(\psi_j, \beta_n^0, r_n) = \left\{f = \sum_{j=1}^{J_n} \beta_j\psi_j \mid \beta = (\beta_1,...,\beta_{J_n})^{\top}\in B_1(\beta_n^0,r_n)\right\},
\end{equation}
with $r_n = r_n(s,D) = \sqrt{\frac{\log J_n}{n}} n^{\frac{1}{2s+1}} \log^{\frac{2s(D-1)}{2s+1}}(n)$. For a specified sequence of $r_n$ and $J_n$, $\mathcal{G}_n$ is a deterministic sequence of function spaces. The set $B_1(\beta,r)\subset \mathbb{R}^{J_n}$ is the $\|\cdot\|_1$-ball of radius $r$ centered at $\beta$, formally
\begin{equation}
   B_1(\beta,r) = \{\gamma \in \mathbb{R}^{J_n}\mid \|\gamma - \beta\|_1 \leq r\} 
\end{equation}

In the rest of this section, we will derive some bounds on the metric entropy of $\mathcal{G}_n$ and apply some maximal inequalities to relate the testing errors to the training errors. We will show that the metric entropy of the function space $\mathcal{G}_n$ (equipped with $\|\cdot\|_n$-norm) is of the same order as the metric entropy of $B_1(\beta_n,r_n)$ (equipped with Euclidean $\|\cdot\|_2$-norm). Since the latter is known in the literature (e.g, Lemma~3 in \cite{raskutti2011minimax}), we have the following results:
\begin{lemma}
\label{lemma:entropyboundonG}
Let $r_n = r_n(s,D) = \sqrt{\frac{\log J_n}{n}} n^{\frac{1}{2s+1}} \log^{\frac{2s(D-1)}{2s+1}}n$. Then for the $\mathcal{G}_n$ defined in \eqref{eq:defgn}, we have
\begin{equation}
    H(\delta,\mathcal{G}_n,P_n)
     \leq cM^{2}r_n^2\delta^{-2}\log J_n
\end{equation}
\end{lemma}
\begin{proof}
We first rewrite the empirical norm in matrix notation:
\begin{equation}
    \|f_{\beta}\|_n = \left\{\frac{1}{n}\sum_{i=1}^nf_{\beta}^2(\mathbf{X}_i)\right\}^{1/2}
    = \frac{1}{\sqrt{n}}\left\{\sum_{i=1}^n\left(\sum_{j=1}^{J_n} \beta_j\psi_j(\mathbf{X}_i)\right)^2\right\}^{1/2} = \left\|\frac{1}{\sqrt{n}}\hat\Psi\beta\right\|_2,  
\end{equation}
where $\hat\Psi$ is the design matrix: $\hat\Psi_{ij} = \psi_j(\mathbf{X}_i)$.

So we know if there is a $\delta$-cover of the set $\left\{\frac{1}{\sqrt{n}}\hat\Psi\beta, \beta \in B_1(\beta_n^0,r_n)\right\}\subset \mathbb{R}^{n}$ under the Euclidean $\|\cdot\|_2$-norm, we can directly construct one for $\mathcal{G}_n$ under $\|\cdot\|_n$. There are available bounds on the covering number of the $\frac{1}{\sqrt{n}}\hat\Psi\beta$ when $\beta$ belongs to a $l_1$-ball. Specifically, we can apply Lemma~4 of \cite{raskutti2011minimax}:
\begin{equation}
    H\left(\delta, \mathcal{G}_n,P_n\right) \leq cr_n^2M^{2}\delta^{-2}\log J_n.
\end{equation}
This concludes our proof.
\end{proof}

To relate the training and testing errors, we need to consider a function space closely related to $\mathcal{G}_n$:
\begin{equation}
    \tilde{\mathcal{G}}_n^2 = (\mathcal{G}_n - f^0)^2
\end{equation}

We summarize several properties of it in the following lemma.
\begin{lemma}
\label{lemma:entropyandother}
Let $r_n = \sqrt{\frac{\log J_n}{n}} n^{\frac{1}{2s+1}} \log^{\frac{2s(D-1)}{2s+1}}n$ and $\delta_n < r_n$. Then for the function space $\tilde{\mathcal{G}}_n^2$ we know:
\begin{equation}
    \begin{aligned}
    & \qquad\sup_{g\in \tilde{\mathcal{G}}_n^2} \|g\|_{\infty} \leq c(M,f^0)\\
        & P\left(\sup_{g\in \tilde{\mathcal{G}}_n^2} \|g\|_n \leq c(M,f^0)r_n\right)\  \stackrel{n \rightarrow \infty}{\longrightarrow} 1 \\
        &\int_{\delta_n}^{r_n} H^{1/2}(u,\tilde{\mathcal{G}}_n^2,P_n) du \leq c(M)r_n (\log J_n)^{1/2}\log(1/\delta_n)
    \end{aligned}
\end{equation}
\end{lemma}

\begin{proof}
\begin{itemize}
    \item We first derive the bound on the $\|\cdot\|_{\infty}$-norm. By definition, every element $g$ in $\tilde{\mathcal{G}}_n^2$ can be expressed as $g = (f - f^0)^2$ for some $f\in \mathcal{G}_n$. To bound $\|g\|_{\infty}$, it is enough to bound $\|f - f^0\|_{\infty}$.
    \begin{equation}
        \|f - f^0\|_{\infty} \leq \|f - f_{\beta_n^0}\|_{\infty} + \|f_{\beta_n^0} - f^0\|_{\infty} \stackrel{(1)}{\leq} Mr_n + \|f_{\beta_n^0}\|_{\infty} + \|f^0\|_{\infty} \leq c(M,s,D, f^0).
    \end{equation}
    In step (1) we used the explicit form of $f$ and $f_{\beta_n^0}$.
    
    \item Bound on the empirical norm $\|\cdot\|_n$. Let $g = f - f^0$ for some $f\in \mathcal{G}_n$. We can bound the empirical norm of it:
    \begin{equation}
    \label{eq:boundonempnorm}
        \begin{aligned}
            \|g\|_n &\leq \|f - f_{\beta_n^0}\|_n + \|f_{\beta_n^0} - f^0\|_n\\
            &\leq Mr_n + c(s,D,\rho_X, M^0)(\log^{D-1}n/n)^{s/2s+1} \text{ with high probability.}
        \end{aligned}
    \end{equation}
    The first term is using the explicit form of $f$ and $f_{\beta_n^0}$. The second bound is based on the approximation results in Lemma~\ref{lemma:approximationsparsemodel} and the probability bound in \eqref{eq:onefunctionratio}. Since $r_n = \sqrt{\frac{\log J_n}{n}} n^{\frac{1}{2s+1}} \log^{\frac{2s(D-1)}{2s+1}}(n)$, the order of the first term in \eqref{eq:boundonempnorm} is larger than the second one's. So we conclude that for $g\in \mathcal{G}_n - f^0$, $\|g\|_n \leq c(M)r_n$. For $g^2\in \tilde{\mathcal{G}}_n^2$,
    \begin{equation}
        \|g^2\|_n^2 = \frac{1}{n}\sum_{i=1}^n \left\{f(\mathbf{X}_i) - f^0(\mathbf{X}_i)\right\}^4 \leq c(M,s, D, M^0)\|g\|_n^2\leq c(M,s,D,M^0)r_n^2.
    \end{equation}
    So we conclude that for any $g^2\in \tilde{\mathcal{G}}_n^2$, $\|g^2\|_n \leq c(M,s,D,M^0)r_n$ with probability $\rightarrow$ 1.
    
    \item Now we derive the bound on the integrated metric entropy. As a first step, we are going to show from a covering of $\mathcal{G}_n$ one can construct one for $\tilde{\mathcal{G}}^2_n$. For any $h_1,h_2 \in \tilde{\mathcal{G}}^2_n$, there exist $f_1,f_2 \in \mathcal{G}_n$ such that $h_i = (f_i - f^0)^2$, $i\in \{1,2\}$. So we know that
    \begin{equation}
        \begin{aligned}
            \|h_1 - h_2\|_n^2 & = \|(f_1 - f^0)^2 - (f_2 - f^0)^2\|_n^2\\
            & = \frac{1}{n}\sum_{i=1}^n \left[ \left\{f_1(\mathbf{X}_i)+f_2 (\mathbf{X}_i)- 2f^0(\mathbf{X}_i)\right\}\left\{f_1(\mathbf{X}_i)-f_2(\mathbf{X}_i)\right\}\right]^2\\
            &\leq c(M,s,D,M^0)\frac{1}{n}\sum_{i=1}^n \left(f_1(\mathbf{X}_i) - f_2(\mathbf{X}_i)\right)^2 = c(M, s,D,M^0)\|f_1 - f_2\|_n^2
        \end{aligned}
    \end{equation}
    
    Now we know that if we have a $\delta$-covering of $\mathcal{G}_n$ with center points $\{f_k\}$, then the functions $\{(f_k - f^0)^2\}$ is a $c(M,s,D,M^0)\delta$-covering of $\tilde{\mathcal{G}}^2_n$. Since we already have an entropy bound on $\mathcal{G}_n$ stated in Lemma~\ref{lemma:entropyboundonG}, we have one for $\tilde{\mathcal{G}}^2_n$ of the same order as well. The integrated entropy can be bounded as follows:
    \begin{equation}
        \begin{aligned}
            \int_{\delta_n}^{r_n} H^{1/2}(\tau,\tilde{\mathcal{G}}^2_n, P_n)d\tau 
            &\leq c(M) \int_{\delta_n}^{r_n} (\log J_n)^{1/2} r_n\tau^{-1}d\tau\\
            &\leq c(M) (\log J_n)^{1/2} r_n\log(1/\delta_n),
        \end{aligned}
    \end{equation}
    when $r_n \leq 1$. 
\end{itemize}
\end{proof}
\begin{theorem}
\label{th:empiricalprocess}
Let 
\begin{equation}
    \begin{aligned}
        &\delta_n = c(s,D)\log (J_n) n^{-\frac{2s}{2s+1}} \log^{\frac{2s(D-1)}{2s+1} + 1}(n),\\
        & r_n = c(s,D)\sqrt{\frac{\log J_n}{n}} n^{\frac{1}{2s+1}} \log^{\frac{2s(D-1)}{2s+1}}(n).
    \end{aligned}
\end{equation}
And let $\hat\beta_n$ denote the minimizer of the penalized problem \eqref{eq:LASSOproblem}. Then, under the same conditions as in Theorem~\ref{th:maintheorem}, we have
\begin{equation}
    \limsup_{n\rightarrow\infty} P\left(\left|\|f_{\hat\beta_n} - f^0\|_n^2 - \|f_{\hat\beta_n} - f^0\|_2^2\right| \geq \delta_n\right)= 0
\end{equation}
\end{theorem}
\begin{proof}
We first need to apply the symmetrization trick (e.g. Corollary 3.4 in \cite{geer2000empirical})
\begin{equation}
\label{eq:aftersymmetrization}
\begin{aligned}
     P\left(\left|\|f_{\hat\beta_n} - f^0\|_n - \|f_{\hat\beta_n} - f^0\|_2\right| \geq \delta_n \right) 
     &= P\left(\left|(P_n-P)\left(f_{\hat\beta_n}-f^0\right)^2\right|\geq \delta_n\right)\\
     &\leq P\left(\sup_{g\in\tilde{\mathcal{G}}_n^2}\left|(P_n-P)g\right|\geq \delta_n\right) + P(f_{\hat\beta_n}\notin \mathcal{G}_n)\\
     &\leq 4P\left(\sup_{g\in\tilde{\mathcal{G}}_n^2}\left|\frac{1}{n}\sum_{i=1}^nW_ig(\mathbf{X}_i)\right|\geq \delta_n/4 \right)+ P(f_{\hat\beta_n}\notin \mathcal{G}_n)
\end{aligned}
\end{equation}
The $W_i$ variables above are independent and identically distributed Rademacher variables ($P(W_i = 1) = P(W_i = -1) = 0.5$). They are bounded (therefore sub-Gaussian) random variables. The probability $P(f_{\hat\beta_n}\notin \mathcal{G}_n)$ has been investigated in Corollary~\ref{corollary:cleanversion2}. To bound the first term in \eqref{eq:aftersymmetrization}, we need to apply some maximal inequalities (e.g., Corollary~8.3 or Lemma~3.2 in \cite{geer2000empirical}). These results require that $r_n > \delta_n$ and
\begin{equation}
        \sqrt{n}\delta_n \geq C\left(\int_{\delta_n/8}^{r_n} H^{1/2}\left(\tau, \tilde{\mathcal{G}}_n^2, P_n\right)d\tau \vee r_n\right).
\end{equation}
We already checked these properties in Lemma~\ref{lemma:entropyandother}. So we conclude that with probability going to $1$, the difference between the training and testing error is no larger than $\delta_n$.
\end{proof}

\renewcommand*{\proofname}{Proof of Theorem~\ref{th:maintheorem}}
\begin{proof}
To show the testing error stated in Theorem~\ref{th:maintheorem}, we just need to combine the results in Theorem~\ref{th:empiricalprocess} and Corollary~\ref{corollary:cleanversion2}.
\end{proof}
\renewcommand*{\proofname}{Proof}

\newpage

\section{The Average Order of Divisor Functions}
\label{section:NumberTheoryResults}
In this section we will derive the average order of $D$-divisor functions that was used in the proof of Lemma~\ref{lemma:numbertheory}. Such a result is well-known in for mathematicians working in the field of number theory, and is usually considered as a direct generalization to the case $D=2$. However, most standard references only include the special (but important) case when $D = 2$. For the purpose of completeness, we replicate a proof based on a public online note by Dr. Graham Jameson \cite{MDIV}. In Theorem MDIV18 of the note, the author also presents the asymptotic constant of the second order term, which is very interesting but less relevant to our purpose, so we choose not to reproduce such a finer result. For other reference of similar and more generalized results, see \cite{huybrechs2011high} (Proposition~6) and \cite{dobrovol1998number}.

\begin{definition}
\label{def:TD}
We define the sequence $T_D(x)$ to be the sum of the $D$-divisor function evaluated at the first $\lfloor x\rfloor$ positive natural numbers, that is
\begin{equation}
T_D(x)=\sum_{n \leq x} \tau_D(n),
\end{equation}
the $D$-divisor function $\tau_D$ is defined in Definition~\ref{def:divisorfunction}.
\end{definition}

Clearly, $T_D(x)$ is the number of $D$-tuples $\mathbf{j} = (\mathbf{j}^1,..,\mathbf{j}^D)\in (\mathbb{N}^+)^D$ with $\prod_{k=1}^{D}\mathbf{j}^{k} \leq x$. We note that $x$ is not necessarily an integer: the summation index $n\leq x$ should be interpreted as $\{1,2,....,\lfloor x \rfloor\}$.

\begin{lemma}
We have the following recurrence relation for $T_D$ (over $D$):
\begin{equation}
\label{eq:resultsonTk}
    T_D(x) =\sum_{n \leq x} T_{D-1}\left(\frac{x}{n}\right) 
\end{equation}
\end{lemma}
\begin{proof}
Fix $n \leq x .$ The number of $D$-tuples $\left(\mathbf{j}^{1}, \mathbf{j}^{2}, \ldots, \mathbf{j}^{D-1}, n\right)$ with $ n\prod_{k=1}^{D-1}\mathbf{j}^k \leq x$ is the number of $(D-1)$-tuples with $\prod_{k=1}^{D-1}\mathbf{j}^k \leq x / n$, that is, $T_{D-1}(x / n)$. Hence $T_D(x)=\sum_{n \leq x} T_{D-1}(x / n)$. 

\end{proof}

\begin{lemma}
Define
\begin{equation}
    A_D(x)=\sum_{n \leq x} \frac{x}{n} \log^D(x/n)
\end{equation}
then we have
 \begin{equation}
\label{eq:14}
    \frac{1}{D+1} x\log ^{D+1} x \leq A_D(x) \leq \frac{1}{D+1} x\log ^{D+1} x +x\log ^D x
\end{equation}
\end{lemma}

\begin{proof}
Let $f(t)=\frac{x}{t}\log^D (x/t)$ for $1 \leq t \leq x$ (also $f(t)=0$ for $t>x$). Then $f(t)$ is decreasing and non-negative, and
\begin{equation}
   \int_{1}^{x} f(t) d t=\left[\frac{x}{D+1} \log ^{D+1} (u)\right]_{1}^{x}=\frac{x\log ^{D+1}(x)}{D+1}
\end{equation}
The statement follows, by using the following basic integral estimation (Proposition 1.4.2 of \cite{jameson2003prime}): Let $f(t)$ be a decreasing, non-negative function for $t \geq 1$. Write $S(x)=\sum_{n \leq x} f(n)$ and $I(x)=\int_{1}^{x} f(t) d t$. Then for all $x \geq 1$,
\begin{equation}
    I(x) \leq S(x) \leq I(x)+f(1)
\end{equation}
\end{proof}

\begin{lemma}
\label{lemma:numbertheory}
For all $D \geq 2$,
\begin{equation}
\label{eq:17}
    T_D(x)=\frac{1}{(D-1) !} x \log ^{D-1} x+O\left(x \log ^{D-2} x\right)
\end{equation}
\end{lemma}

\begin{proof}
Induction on $D$. The case $D=2$ is known to be true (Theorem~3.2 \cite{tenenbaum2015introduction}). Assume \eqref{eq:17} for $D$, with the error term denoted by $q_D(x)$. Then by $\eqref{eq:resultsonTk}$,
\begin{equation}
    T_{D+1}(x)=\sum_{n \leq x} T_D\left(\frac{x}{n}\right)=I(x)+Q(x)
\end{equation}
where
\begin{equation}
    \begin{aligned}
        I(x)=\frac{1}{(D-1) !} \sum_{n \leq x} \frac{x}{n} \log ^{D-1} \frac{x}{n}=\frac{1}{(D-1) !} A_{D-1}(x) \\
Q(x)=\sum_{n \leq x}  q_D\left(\frac{x}{n}\right) \sim \sum_{n \leq x}  \frac{x}{n}\log ^{D-2} \frac{x}{n}=A_{D-2}(x)
    \end{aligned}
\end{equation}
By \eqref{eq:14},
\begin{equation}
    I(x)=\frac{1}{D !} x\log ^{D} x+O\left(x\log ^{D-1} x\right)
\end{equation}
and $Q(x) = O(x\log ^{D-1} x)$. Hence \eqref{eq:17} holds for $D+1$. 
\end{proof}

\newpage

\bibliographystyle{apalike}
\bibliography{main}
\end{document}